\documentclass[preprint,12pt]{elsarticle}

\usepackage{amssymb}
\usepackage{graphics}
\usepackage{amsmath}
\usepackage{color}

\journal{Journal of Molecular Liquids}

\begin{document}

\begin{frontmatter}



\title{Phase behaviour in ionic solutions: restricted primitive model of ionic liquid in explicit neutral solvent}


\author{O.V. Patsahan,  T.M. Patsahan}

\address{Institute for Condensed Matter Physics of the National
Academy of Sciences of Ukraine, 1 Svientsitskii St., 79011 Lviv,
Ukraine}

\begin{abstract}
We study  fluid-fluid equilibrium in the simplest model of ionic solutions where the solvent is  explicitly included, i.e., 
a binary mixture consisting of
a restricted primitive model (RPM) and neutral hard-spheres (RPM-HS mixture).
 First, using the collective variable method we find  free energy, pressure and  partial chemical potentials in the random phase approximation (RPA) for a rather general model that takes into consideration solvent-solvent and solvent-ion interactions beyond the hard core.
 In the special case of a RPM-HS mixture,
we consider two  regularizations  of the Coulomb potential
inside the hard core, i.e.,  the Weeks-Chandler-Andersen (WCA) regularization leading to the WCA approximation and  the optimized 
regularization giving the optimized RPA (ORPA) or 
the mean spherical approximation (MSA). Furthermore,  we calculate the phase coexistence using  the associative mean spherical approximation (AMSA). 
In general, the three approximations produce qualitatively 
similar phase diagrams of the RPM-HS mixture, i.e.,  a fluid-fluid coexistence envelope with an upper critical solution point,   a shift of  the coexistence region towards higher total number densities  and higher solvent 
concentrations with increasing pressure, and  a  small increase of the  critical temperature with an increase of pressure. As for a pure RPM, 
the AMSA  leads to the best agreement 
with the available simulation data when the  association constant  proposed by Olaussen and Stell is used. We also discuss the peculiarities of the phase diagrams  in the WCA approximation.  

\end{abstract}

\begin{keyword}
ionic liquids \sep liquid-liquid equilibrium \sep explicit solvent model\sep collective variable approach \sep associative mean spherical approximation
\PACS 05.20.-y \sep 64.60.De \sep 64.75.Cd \sep 64.75.Gh \sep 82.60.Lf        


\end{keyword}

\end{frontmatter}


\section{Introduction}
The liquid-liquid phase
equilibrium in electrolyte solutions and room temperature ionic liquids (ILs) represents one of the most important problems
in physical chemistry and chemical engineering. In particular, the understanding of
phase behaviour is essential in extraction, purification and many other industrial applications.

Most theoretical and computer simulation studies of ionic systems are based on the restricted primitive  model (RPM),
i.e., an equimolar mixture of
equisized  charged hard spheres immersed in a structureless
dielectric continuum.
Theoretical studies of the RPM
predict the vapour-liquid-like phase transition at low reduced temperatures and at low reduced densities
\cite{stell1,levin-fisher,Cai-Mol1,patsahan_ion}.  Reliable estimates
of  the  location of the critical point have been obtained by using mixed-field finite-size (MFFS) scaling methods of computer 
simulations \cite{caillol-levesque-weis:02,Hynnien-Panagiotopoulos}.
In accordance
 with the prediction of the RPM, the liquid-liquid critical points of ILs in non-polar solvents are located at low temperatures
 and at low concentrations when the critical parameters are expressed in terms of the RPM variables 
 \cite{Saracsan,Wagner:03,Butka:08,Schroer_Vale:09}. 
However,
systematic deviations of the reduced
critical temperature with the dielectric constant of a solvent indicate the limitations
of the analogy of a liquid-liquid transition in ionic solutions compared to the
vapour-liquid transition of the RPM \cite{Butka:08,Schroer_Vale:09,Schroer:12}.
Modified versions of the RPM that take into account the charge or/and size
asymmetry, the so-called primitive models (PM), were also studied theoretically
\cite{Gonzalez-Tovar,Kalyuzhnyj-Holovko-Vlachy:00,zuckerman:01,artyomov:03,fisher_aqua_banerjee,patsahan-mryglod-patsahan:06,PatPat09,Patsahan_Patsahan:10,PatMryg12} 
and by Monte Carlo (MC)  simulations 
\cite{Camp-Patay:99,Romero-Enrique:00,
yan-pablo:01,yan-pablo:02,yan-pablo:02:2,panagiot-fisher:02,panagiotopoulos1,cheong-panagiot:03,kim-fisher-panagiotopoulos:05}. 
A comprehensive overview of the phase diagrams of ILs in a variety of polar and non-polar solvents  and the comparison of 
these diagrams with the results of model systems, in particular, the PM and the RPM, show that more complex models should be 
used to properly 
describe these systems \cite{Schroer_Vale:09,Schroer:12}.   In this connection,
models that take  the  structure of solvents into consideration are of particular interest.

In the simplest model that takes into account a discrete structure of the solvent,
the ions are modelled by charged hard spheres
while the solvent molecules are neutral hard spheres.  The polar
nature of the solvent is represented implicitly by a continuum
background with a dielectric constant.  This model is often called a solvent primitive model or   the SPM (see \cite{Kristof_SPM} and references therein).
However, the above-mentioned abbreviation was also used for another model, the so-called special primitive model, 
that is the PM with the same ion diameter but with different valences (see e.g. \cite{Cai-Mol1}). To avoid misunderstandings, thereafter we  refer to  a binary mixture of the PM and the neutral  hard spheres as the PM-HS mixture (or the RPM-HS mixture for the equisized monovalent PM).   
The phase behaviour of the RPM-HS mixture was  theoretically studied by using the mean-spherical approximation (MSA) \cite{Kenkare_SPM} and the 
pairing MSA (PMSA) \cite{Zhou_SPM}. For the pure RPM fluid, the PMSA theory  leads only to a slight decrease 
of the critical temperature  when compared to the MSA results but it noticeably improves the MSA critical density \cite{Zhou_RPM}.
As it was shown  in \cite{Kenkare_SPM,Zhou_SPM},  the critical temperature of the RPM-HS mixture expressed in the RPM reduced variables 
is a bit  higher than for the RPM.
 To the best of our knowledge, only  a few MC simulation results 
of phase coexistence in the RPM-HS fluid have been reported \cite{Kristof_SPM,Shelley_SPM}.   They were obtained without the use of the MFFS scaling.

The purpose of this paper is to study the phase behaviour  of ionic models 
that take into account the presence of the solvent explicitly. To this end, we use two theoretical approaches, i.e.,
the theory that uses the collective variable (CV) method 
\cite{Pat-Mryg-CM,Patsahan_Mryglod_Caillol} and the associative mean spherical approximation (AMSA) that exploits the concept 
of ion association \cite{HolKalyuzh91,Blum95,Bernard96}. For the RPM-HS mixture, the AMSA theory reduces 
to the MSA when the association between ions is neglected. On the other hand, the Helmholtz free energy in the MSA can be
obtained within the framework of the CV  theory from the random phase approximation (RPA) using an optimized 
regularization  of the Coulomb potential inside  the hard core \cite{anderson_chandler}.

By applying the CV theory, we start with a  general model where  the interaction potentials  
between the two solvent particles
and between the ion and the solvent particle  include a short-range attraction/repulsion in
addition to  a hard-core repulsion. Moreover, the positive and negative ions and the ions and the solvent particles can differ in size. 
For this model,  free energy in the RPA is obtained. Here, we address the phase diagrams
of an equisized RPM-HS mixture.  We calculate the phase diagrams in the RPA  using the Weeks-Chandler-Andersen (WCA) 
regularization  of the Coulomb potential inside the hard core \cite{cha,wcha}. Similar to \cite{Cai-Mol1}, we refer to this approximation as the WCA approximation.  Further, we calculate the phase diagrams of the model 
in the  MSA and the AMSA.  We compare the results  obtained in the above-mentioned approximations.  In addition, the results for the 
RPM-HS  are compared with those of the RPM. The three  approaches  yield
the results for the  liquid-liquid phase coexistence of the RPM-HS mixture that are in qualitative agreement.  However, the best agreement 
with the available simulation findings is achieved  for the case of the AMSA.
At the same time,  we get  the lower critical solution temperature (LCST)-type of phase transition and  
the evidence of a closed miscibility loop in the WCA approximation. By contrast, the lower critical solution point (LCSP) 
is  found neither in the MSA nor in the AMSA approaches. Using the thermodynamic relations we show that the obtained LCSPs are not stable 
critical points. 

The paper is arranged as follows. In section~2, we present two theoretical formalisms, i.e., the CV theory and the AMSA theory.
In Section~3, the phase diagrams of the RPM-HS mixture  are calculated in the RPA, the MSA, and the AMSA. Here, an  analysis of the
results and their discussion are presented.  
We draw conclusions in section 4.

 \section{Theoretical formalism}
 \subsection{Method of collective variables}

We consider  the primitive model of ionic fluids  consisting of $N_{+}$ hard spheres  of diameter
$\sigma_{+}$  carrying a charge $q_{+}$ and $N_{-}$ hard spheres of
diameter $\sigma_{-}$ carrying a charge $q_{-}$ ($q_{+}N_{+} +q_{-} N_{-} =0$).
The ionic model is  immersed in a solvent consisting of  $N_{s}$ uncharged (neutral) particles.
The pair interaction potentials are assumed to be of the following form:
\begin{equation}
U_{\alpha \beta } (r)=\phi _{\alpha \beta }^{\rm RS} (r)+\phi _{\alpha \beta }(r),
\label{2.1}
\end{equation}
where indices  $\alpha ,\beta =s,+,-$ denote the corresponding species.  In (\ref{2.1}),
 $\phi _{\alpha \beta }^{RS} (r)$ is the potential of a short-range repulsion which describes the mutual impenetrability
 of the particles and  $\phi _{\alpha \beta}(r)$  describes the behaviour at moderate and large distances.
 In our case, $\phi_{\alpha\beta}^{RS}(r)$ is the interaction potential between the two  additive hard spheres of diameters
$\sigma_{\alpha}$ and $\sigma_{\beta}$.
 Thermodynamic  and structural properties of the system  interacting via the potential $\phi _{\alpha \beta }^{RS} (r)$
are assumed to be known and, therefore, this system can be regarded as a reference system (RS).
In the case of ions,  $\phi _{\alpha \beta}(r)$ is the Coulomb potential $\phi_{\alpha\beta}(r)=q_{\alpha}q_{\beta}\phi^{C}(r)$,
$\phi^{C}(r)=1/(\varepsilon r)$, $\varepsilon$ is the dielectric constant of the solvent. At this stage, we do not specify the form
of the interaction potentials $\phi_{\alpha\beta}(r)$ acting between two neutral particles and between charged and neutral particles.

 The  model (\ref{2.1}) is  at equilibrium in the grand
canonical ensemble. Then, the
grand partition function of the model  reads
\begin{eqnarray*}
\Xi=\sum_{N_{+}\geq 0}\sum_{N_{-}\geq
0}\sum_{N_{s}\geq 0}\;\prod_{\alpha=+,-,s}
\frac{\exp(\nu_{\alpha}N_{\alpha})}{N_{\alpha}!} \int({\rm d}\Gamma)
\exp\left[-\frac{\beta}{2}\sum_{\alpha\beta}\sum_{ij}
U_{\alpha\beta}(r_{ij})\right],
\end{eqnarray*}
where  $\nu_{\alpha}$ is the
dimensionless chemical potential,
$\nu_{\alpha}=\beta\mu_{\alpha}-3\ln\Lambda_{\alpha}$,
$\mu_{\alpha}$ is the chemical potential of the $\alpha$th species,
$\beta$ is the reciprocal temperature, $\Lambda_{\alpha}^{-1}=(2\pi
m_{\alpha}\beta^{-1}/h^{2})^{1/2}$ is the inverse de Broglie thermal
wavelength; $(\rm d\Gamma)$ is the element of configurational space
of the particles.

Using a formalism of the collective variable (CV) method, we can present the functional of the grand partition function of the
above-described model in the form \cite{Pat-Mryg-CM,Patsahan_Mryglod_Caillol}:
\begin{eqnarray}
&&\Xi =\Xi _{{\rm MF}} \int ({\rm d}\rho )({\rm d}\omega )\exp\left
[-\frac{\beta }{2V} \sum _{\alpha ,\beta } \sum _{\mathbf k} \tilde{\phi}_{\alpha \beta
} (k)\rho _{\mathbf k,\alpha } \rho _{-\mathbf k,\beta } \right. \nonumber \\
&&
\left.
+{\rm i}\sum _{\alpha} \sum _{\mathbf k} \omega
_{{\mathbf k},\alpha } \rho _{{\mathbf k},\alpha }
-\frac{1}{2} \sum _{\alpha ,\beta } \sum _{\mathbf k} {\mathfrak M}_{
\alpha \beta } (k)\omega _{{\mathbf k},\alpha } \omega _{-{\mathbf k},\beta } +\sum _{n\ge 3}\frac{(-{
\rm i})^{n} }{n!} \delta H_{n}\right].
\label{Xi_CV}
\end{eqnarray}
In (\ref{Xi_CV}), the following notations are introduced. $\Xi _{MF} $ is the mean-field (MF) part of the grand partition function:
\begin{eqnarray*}
 \ln\Xi_{MF} =\ln\Xi_{RS} +\frac{\beta}{2}\left[\tilde{\phi}_{ss}
\bar{\rho}_{s}^{2} +2\tilde{\phi}_{s+}\bar{\rho}_{s} \bar{\rho}_{+} +2
\tilde{\phi}_{s-}\bar{\rho}_{s}\bar{\rho}_{-}\right],
\end{eqnarray*}
$\Xi_{\rm{RS}}$ is the grand partition function  of a
two-component hard sphere system, $\bar{\rho }_{\alpha } =\left\langle N_{
\alpha }/V \right\rangle _{RS} $, $\langle\ldots\rangle_{_{RS}}$
indicates the  average taken over the RS.

$\rho_{{\mathbf k},\alpha}=\rho_{{\mathbf k},\alpha}^c-{\rm
i}\rho_{{\mathbf k},\alpha}^s$  is the  CV that describes the value
of the $\mathbf k$-th fluctuation mode of the number density of the
$\alpha$th species, the indices $c$ and $s$ denote real and
imaginary parts of $\rho_{{\mathbf k},\alpha}$.  $\omega_{{\mathbf
k},\alpha}$ is conjugate to the CV $\rho_{{\mathbf k},\alpha}$ and
each of $\rho_{{\mathbf k},\alpha}$ ($\omega_{{\mathbf k},\alpha}$)
takes all the real values from $-\infty$ to $+\infty$. $({\rm
d}\rho)$  and $({\rm d}\omega)$ are  volume elements of the CV phase
space
\begin{displaymath}
({\rm d}\rho)=\prod_{\alpha}{\rm d}\rho_{0,\alpha}{\prod_{\mathbf
k\not=0}}{\rm d}\rho_{\mathbf k,\alpha}^{c}{\rm d}\rho_{\mathbf
k,\alpha}^{s}, \quad ({\rm d}\omega)=\prod_{\alpha}{\rm
d}\omega_{0,\alpha}{\prod_{\mathbf k\not=0}}{\rm d}\omega_{\mathbf
k,\alpha}^{c}{\rm d}\omega_{\mathbf k,\alpha}^{s}
\end{displaymath}
and the product over ${\mathbf k}$ is performed in the upper
semi-space ($\rho_{-\mathbf k,\alpha}=\rho_{\mathbf k,\alpha}^{*}$, $\omega_{-\mathbf k,\alpha}=\omega_{\mathbf k,\alpha}^{*}$).
$\tilde \phi_{\alpha\beta}(k)$ is the Fourier
transform  of the interaction potential $\phi_{\alpha\beta}(r)$.
 
For $\delta H_{n}$, we have:
\begin{equation*}
 \delta H_{n} =\sum _{\alpha _{1},\ldots,\alpha _{n} } \sum _{{\mathbf k}_{1},\ldots,{\mathbf k}_{n} } {\rm
{\mathfrak M}}_{\alpha _{1},\ldots,\alpha _{n} } ({\mathbf k}_{1} ,\ldots ,{\mathbf k}_{n} )\omega _{{\mathbf k}_{1}
,\alpha _{1} } \ldots \omega _{{\mathbf k}_{n},\alpha _{n} } \delta _{{\mathbf k}_{1} +\ldots+{\mathbf k}_{n}},
\end{equation*}
where the $n$th cumulant ${\mathfrak{M}}_{\alpha_{1}\ldots\alpha_{n}}$
coincides with the Fourier transform of the $n$-particle connected correlation function of the RS \cite{Pat-Mryg-CM},
$\delta_{{\bf{k}}_{1}+\ldots+{\bf{k}}_{n}}$ is the Kronecker symbol. It should be noted that ${\mathfrak{M}}_{\alpha_{1}\ldots\alpha_{n}}$
depends on the renormalized chemical potential $\bar\nu_{\alpha}=\nu_{\alpha}+\nu_{\alpha}^{s}$, where
$\nu_{\alpha}^{s}=\frac{\beta}{2V}\sum_{{\mathbf k}}\tilde\phi_{\alpha\alpha}(k)$ is the self-energy of the
$\alpha$th species \cite{Pat-Mryg-CM}.

\paragraph{Random phase approximation}
Now, we restrict  our consideration to the second-order cumulants ${\mathfrak M}_{\alpha \beta} (k)$ setting  $\delta H_{n}=0$ in (\ref{Xi_CV}). Then, after
integration in (\ref{Xi_CV}) over  CVs  $\omega_{{\mathbf{k}},\alpha}$ and  $\rho _{{\mathbf{k}},\alpha}$ we get the grand
partition function in the Gaussian approximation
\begin{eqnarray*}
 \ln\Xi _{G} =\ln \Xi _{MF} -\frac{1}{2} \sum _{{\mathbf k}}
\ln \det \left[\underline{1}+\hat{\Phi }_{2} \hat{{\mathfrak M}}_{2} \right],
\end{eqnarray*}
where $\hat{\Phi }_{2}$ and $\hat{\mathfrak M}_{2}$
denote  symmetric $3\times3$  matrices of elements
$\beta\tilde\phi_{\alpha\beta}(k)$ and
${\mathfrak{M}}_{\alpha\beta}(k)$, respectively, $\underline{1}$ is the unit matrix.

Following the Legendre transform of $\ln\Xi _{G}$ we arrive at the Helmholtz free energy per volume,
$\beta f=\beta F/V$,  in the random phase approximation (RPA)
\begin{eqnarray}
&& \beta f_{RPA} =\beta f^{RS} +\frac{\beta }{2} \left[\tilde{\phi }_{ss}(0)
\rho _{s}^{2} +2\tilde{\phi}_{s+}(0)\rho _{s} \rho _{+} +2\tilde{
\phi }_{s-}(0)\rho _{s} \rho _{-} \right] \nonumber \\
&&
-\frac{\beta }{2V}
\sum _{\alpha =s,+,-}\sum _{{\mathbf k}}\tilde{\phi }_{\alpha \alpha } (k) \rho _{\alpha }
+\frac{1}{2V} \sum _{{\mathbf k}}\ln \det [\underline{1}+\hat{\Phi }_{2} \hat{{\mathfrak M}}_{2}
],
\label{f_RPA}
\end{eqnarray}
where $f^{RS}$ is the free energy of the RS
\begin{equation*}
\beta f^{RS}=\beta f^{ID}+\beta f^{HS},
\end{equation*}
$f^{ID}$ and $f^{HS}$ are the contributions from  ideal gas and hard sphere subsystems, respectively.
The two-particle cumulant ${\mathfrak{M}}_{\alpha\beta}(k)$ is of the form:
\begin{equation*}
 {\mathfrak M}_{\alpha \beta }(k)=\rho _{\alpha }\delta _{\alpha \beta }
+\rho _{\alpha }\rho _{\beta }\tilde{h}_{\alpha\beta}^{RS}(k),
\end{equation*}
where $\rho _{\alpha}$ is the number density of the $\alpha$th species and $\tilde{h}_{\alpha \beta }^{RS} (k)$ denotes the
Fourier transform of the two-particle correlation function  of the RS.

Now, we consider a symmetrical version of the  PM ($q_{+}=|q_{-}|=q $,  $\sigma _{+} =\sigma _{-} =\sigma _{i}$),
the so-called restricted primitive model (RPM). We also assume $\phi _{s+}(r)=\phi _{s-}(r)=\phi _{si}(r)$
(equal solvent-ion interactions).
In this case, we have ${\mathfrak M}_{++}
={\mathfrak M}_{--} $ and ${\mathfrak M}_{s+} ={\mathfrak M}_{s-}={\mathfrak M}_{si}$. As a result, Eq.~(\ref{f_RPA})
takes the form:
\begin{eqnarray}
&& \beta f_{RPA} =\beta f_{RS} +\frac{\beta
}{2}\left[\tilde{\phi }_{ss}(0)\rho _{s}^{2} +2\tilde{\phi }_{si}(0)
\rho _{s} \rho _{i}\right]
-\frac{\beta}{2V}\sum _{\alpha
=s,+,-}\sum _{{\mathbf k}}\tilde{\phi }_{\alpha \alpha }(k)\rho _{\alpha }\nonumber \\
&&
+\frac{1}{2V}
\sum _{{\mathbf k}}\ln [1+\rho _{i} \beta q^{2}\tilde{\phi}^{C}(k)]
+\frac{1}{2V}\sum
_{k}\ln [1+4\beta \tilde{\phi}_{si} (k){\mathfrak M}_{si} +\beta \tilde{\phi
}_{ss} (k){\mathfrak M}_{ss} \nonumber \\
&&
+\beta ^{2} \tilde{\phi}_{si}^{2}(k)(4{\mathfrak M}_{si}^{2} -{\mathfrak M}_{ss}{\mathfrak M}_{ii})],
\label{f_1}
\end{eqnarray}

Using (\ref{f_1}), one can get expressions for the reduced chemical potentials $\nu
_{i} =\nu _{+}=\nu _{-}$ and $\nu _{s}$ in the RPA
\begin{equation}
 \nu _{i}^{RPA} =\nu _{i}^{RS} +\beta \tilde{\phi }_{si} (0)\rho _{s}
-\frac{\beta }{2V}\sum_{{\mathbf k}}q^{2}\tilde{\phi }^{C} (k)+\frac{1}{2V} \sum_{{\mathbf k}}\tilde{g}(k),
\label{nu_i}
\end{equation}
\begin{equation}
 \nu_{s}^{RPA} =\nu _{s}^{RS} +\beta \tilde{\phi }_{ss} (0)\rho _{s}
+\beta \tilde{\phi }_{si} (0)\rho _{i} -\frac{\beta }{2V} \sum_{{\mathbf k}}\tilde{\phi }_{ss}
(k),
\label{nu_s}
\end{equation}
where  $\nu _{i}^{RS} $ and $\nu _{s}^{RS}$ are the RS parts of the corresponding chemical potentials and $\tilde{g}(k)$
is the Fourier transform of the screened Coulomb potential
\begin{displaymath}
 \tilde{g}(k)=\frac{\beta q^{2}\tilde{\phi }^{C} (k)}{1+\beta q^{2}\tilde{\phi }^{C} (k)\rho
_{i}}.
\end{displaymath}
Taking into account Eqs. (\ref{f_1})-(\ref{nu_s}), an expression for pressure $\beta P=\sum _{\alpha }\rho _{\alpha
} \nu _{\alpha } -\beta f$ in the RPA is as follows:
\begin{eqnarray}
 \beta P_{RPA}& =&\beta P^{RS} +\frac{\beta }{2} \left[\tilde{\phi }_{ss} (0)\rho _{s}^{2} +2\tilde{\phi }_{si}
(0)\rho _{s} \rho _{i} \right]
+\frac{\rho _{i} }{2V} \sum _{{\mathbf k}}\tilde{g}(k) \nonumber
\\
&&
-\frac{1}{2V} \sum _{{\mathbf k}}\ln [1+\rho _{i} \beta \tilde{\phi }^{C} (k)]
-\frac{1}{2V}
\sum _{{\mathbf k}}\ln [1+4\beta \tilde{\phi }_{si} (k){\mathfrak M}_{si} \nonumber\\
&&
+\beta \tilde{
\phi }_{ss} (k){\mathfrak M}_{ss} +\beta ^{2} \tilde{\phi }_{si}^{2} (k)(4
{\mathfrak M}_{si}^{2} -{\mathfrak M}_{ss}{\mathfrak M}_{ii})].
\label{P_RPA}
\end{eqnarray}
If the interactions between  solvent particles and between solvent and charged particles can be neglected beyond the hard core, our system
reduces to the RPM-HS mixture.  In this case, from  (\ref{f_1}) we obtain 
 \begin{eqnarray}
\beta f_{RPA} =\beta f^{RS} -\frac{1}{2V}\sum _{{\mathbf k}}\beta q^{2}\tilde{
\phi}^{C}(k)\rho _{i}
+
\frac{1}{2V} \sum _{{\mathbf k}}\ln[1+\rho _{i}\beta q^{2}\tilde{\phi}^{C}(k)].
\label{f_2}
\end{eqnarray}
For $\sigma_{i}=\sigma_{s}=\sigma$   corresponding to a one-component RS,  
Eq.~ (\ref{f_2})  leads to the following expressions for the partial chemical potentials and pressure in the RPA:
\begin{eqnarray}
 \nu_{i}^{RPA}&=&\ln\rho_{i}-\ln 2 +\frac{\eta(8-9\eta+3\eta^{2})}{(1-\eta)^{3}} \nonumber 
 \\
 &&
  -\frac{\beta }{2V}\sum_{{\mathbf k}}q^{2}\tilde{\phi }^{C} (k)
 +\frac{1}{2V} \sum_{{\mathbf k}}\tilde{g}(k),
 \label{nui_RPA_1}
 \\
 \nu_{s}^{RPA}&=& \ln\rho_{s}-\ln 2 +\frac{\eta(8-9\eta+3\eta^{2})}{(1-\eta)^{3}},
 \label{nus_RPA_1}
 \\
 \beta P^{RPA}&=&\frac{\rho(1+\eta+\eta^{2}-\eta^{3})}{(1-\eta)^{3}}+\frac{\rho _{i} }{2V} \sum _{{\mathbf k}}\tilde{g}(k) \nonumber
 \\
 &&
-\frac{1}{2V} \sum _{{\mathbf k}}\ln [1+\rho _{i} \beta \tilde{\phi }^{C} (k)],
\label{P_RPA_1}
\end{eqnarray}
where $\rho=\rho_{i}+\rho_{s}$ is the total number density of an ionic solution and $\eta=\frac{\pi\rho\sigma^{3}}{6}$ is the total 
packing fraction. In (\ref{nui_RPA_1})-(\ref{P_RPA_1}), the Carnahan-Starling (CS) \cite{carnahan-starling} approximation is used 
for the hard-sphere system.

Hereafter, we focus on the  RPM-HS mixture.
First, however,  some comments are in order regarding the form of the potential
$\phi_{\alpha\beta}^{C}(r)$ inside the hard core. The regularization of $\phi_{\alpha\beta}^{C}(r)$
in the physically inaccessible region is  somewhat arbitrary 
 and different regularization schemes for the Coulomb
potential were proposed and used earlier,  see for example \cite{Cai-Mol1,patsahan_ion,anderson_chandler,wcha,waisman_lebowitz,ciach:00:0}.
Within the
framework of the RPA, the best estimation
for the critical temperature of the RPM was achieved for the optimized
regularization \cite{waisman_lebowitz} which leads to the optimized RPA (ORPA). The ORPA
is equivalent to the mean spherical approximation (MSA) where the reference
system is approximated by the Percus-Yevick theory. In this case  \cite{anderson_chandler},
\begin{equation}
\phi^{C}(r) = \left\{
                     \begin{array}{ll}
                     \frac{B}{\varepsilon\sigma_{i}}\left(2-\frac{Br}{\sigma_{i}}\right), & r<\sigma_{i}\\

                     \frac{1}{\varepsilon r},& r\geqslant \sigma_{i}
                     \end{array}
              \right.,
\label{reg-MSA}
\end{equation}
where
\begin{equation}
B=\frac{x^{2}+x-x(1+2x)^{1/2}}{x^{2}}, \qquad x=\kappa_{D}\sigma_{i},
\label{reg-MSA1}
\end{equation}
and $\kappa_{D}^{2}=4\pi q^{2}\rho_{i}/(\varepsilon k_{B}T)$ is the inverse squared Debye length.
The Fourier transform of (\ref{reg-MSA}) is of the form:
\begin{eqnarray}
 \tilde{\phi}^{C}(y)=\frac{4\pi\sigma_{i}^{2}}{\varepsilon y^{4}}\left[2B(1-B)y\sin y+(1-B^{2})y^{2}\cos y-2B^{2}(\cos y-1)\right],
 \label{reg-MSA2}
\end{eqnarray}
where $y=k\sigma_{i}$.
Using the above regularization,  from (\ref{f_2}) one can obtain the well known result for the electrostatic part of the  ORPA free energy
\cite{Cai-Mol1,anderson_chandler,waisman_lebowitz}
\begin{eqnarray*}
 \beta f_{ORPA}-\beta f^{RS}= -\frac{1}{12\pi\sigma_{i}^{3}}\left[6x+3x^{2}+2-2(1+2x)^{3/2}\right].
\end{eqnarray*}

Another choice of the Coulomb potential inside the hard core, also known as the Weeks-Chandler-Andersen (WCA) regularization,
was proposed in \cite{cha}. According to the WCA scheme,
\begin{equation}
\phi^{C}(r) = \left\{
                     \begin{array}{ll}
                 \frac{1}{\varepsilon\sigma_{i}}, & r<\sigma_{i}\\
                 \frac{1}{\varepsilon r},& r\geqslant \sigma_{i}
                     \end{array}
              \right.
\label{reg-WCA}
\end{equation}
and the Fourier transform $\tilde\phi^{C}(y)$ is of the form:
\begin{equation}
 \tilde\phi^{C}(y)=4\pi\sigma_{i}^{2}\frac{\sin y}{\epsilon
y^{3}}.
\label{wca_coul}
\end{equation}
It is worth noting that the regularization (\ref{reg-WCA}) provides rapid convergence of the series of the perturbation theory 
for the free energy \cite{cha}. 

\subsection{Associative mean spherical approximation }

 The associative mean-spherical approximation (AMSA) theory \cite{HolKalyuzh91,Hol05} is based on the modern theory of associating fluids
\cite{Werth84,Wertheim84,Hol02}. As it was shown \cite{Jiang02},  the AMSA  provides much better predictions for the vapour-liquid critical 
parameters of the RPM than the  MSA. %
It is worth noting that 
the AMSA represents the two-density version of the traditional MSA  theory \cite{waisman_lebowitz,Blum75} for an ionic fluid of associative particles.

For the RPM-HS mixture  ($\sigma_{+}=\sigma_{-}=\sigma_{i}$),  the AMSA free energy  can be presented as a sum of three contributions 
\cite{Hol05}:
\begin{equation*}
\beta f_{\rm{AMSA}}=\beta f^{\rm{ID}}+\beta f^{\rm{HS}}+\beta f^{\rm{MAL}}+\beta f^{\rm{EL}},
\end{equation*}
 where    $f^{\rm{ID}}$ and $f^{\rm{HS}}$ are the contributions from the ideal gas and hard sphere 
 subsystems, respectively. These contributions  coincide with the corresponding addends in Eq.~(\ref{f_2}). 
 $f^{\rm{MAL}}$ is the contribution from the mass action law (MAL):
\begin{equation}
\beta f^{\rm{MAL}}=\rho_{i}\ln\alpha+\frac{\rho_{i}}{2}\left(1-\alpha\right),
\label{f_mal}
\end{equation}
where $\alpha$ is the degree of dissociation and according to the MAL it can be found from the following expression \cite{Hol05,Jiang02}:
\begin{equation}
1-\alpha=\rho_{i}\alpha^{2}K,
\label{mal}
\end{equation}
$K=K^{\gamma}K^{0}$ is the association constant, $K^{0}$ is the equilibrium constant of the formation of ion pairs (the so-called thermodynamic association constant), and $K^{\gamma}$ is given by
\begin{equation*}
K^{\gamma}=g_{+-}^{hs}(\sigma_{i})\exp\left[-b\frac{\Gamma^{B}\sigma_{i} (2+\Gamma^{B}\sigma_{i})}
{(1+\Gamma^{B}\sigma_{i})^{2}}\right],
\end{equation*}
where $b$ is the dimensionless Bjerrum length,  $\Gamma^{B}$ is the screening parameter calculated from the equation \cite{Blum95,Bernard96}
\begin{equation}
4\left(\Gamma^{B}\right)^{2}\left(1+\Gamma^{B}\sigma_{i}\right)^{3}=\kappa_{D}^{2}\left(\alpha+\Gamma^{B}\sigma_{i}\right).
\label{Gamma_B}
\end{equation}
It should be noted that without association ($\alpha=1$), $\Gamma^{B}$
reduces to the screening parameter in the MSA \cite{Waisman72,WaismanLeb72,Blum74,Blum75}
\begin{equation*}
\Gamma\sigma_{i}=\frac12\left[\sqrt{1+2\kappa_{D}\sigma_i}-1\right]=\frac{x}{2}(1-B),
\end{equation*}
where $B$ and $x$ are given in (\ref{reg-MSA1}). 

$g_{+-}^{hs}(\sigma_{i})$ is the contact value of the
radial distribution function between the hard-spheres of diameter $\sigma_{i}$. In the Carnahan-Starling (CS) approximation,
for    $g_{+-}^{hs}(\sigma_{i})$ we have \cite{Solana}
\begin{equation*}
g_{+-}^{hs}(\sigma_{i})=\frac{1}{1-\eta}+\frac{3}{2}\frac{\eta}{(1-\eta)^{2}}+\frac{1}{2}\frac{\eta^{2}}{(1-\eta)^{3}},
\end{equation*}
$\eta=\eta_{i}+\eta_{s}$ is the total packing fraction.

$f^{\rm{EL}}$ is the contribution from the electrostatic ion interactions. In   the simple interpolation scheme  approximation 
introduced by Stell and Zhou \cite{Stell89}, this contribution reads
\begin{equation}
\beta f^{\rm{EL}}=-\frac{\beta q^{2}}{\varepsilon}\rho_{i}\frac{\Gamma}{1+\Gamma\sigma_{i}} +\frac{\left(\Gamma \right)^{3}}{3\pi}.
\label{f_el}
\end{equation}

From Eqs. (\ref{f_mal}) and (\ref{mal}), one obtains the following expressions for $P^{\rm{MAL}}$ and $\nu_{i}^{\rm{MAL}}$:
\begin{eqnarray}
\beta P^{\rm{MAL}}&=&-\frac{\rho_i}{2}(1-\alpha)\left(1+\rho_i\frac{\partial\ln K^{\gamma}}{\partial\rho_i}\right),
\label{P_mal} \\
\nu^{\rm{MAL}}_{i}&=&\ln\alpha-\frac{\rho_i}{2}(1-\alpha)\frac{\partial\ln K^{\gamma}}{\partial\rho_i}.
\label{mu_mal}
\end{eqnarray}

Accordingly, $P^{\rm{EL}}$ and $\nu_{i}^{\rm{EL}}$ can be found from Eq.~(\ref{f_el})
\begin{equation}
\beta P^{\rm{EL}}=-\frac{\Gamma^{3}}{3\pi}, \qquad
\nu_{i}^{\rm{EL}}=-\frac{1}{T^{*}}\frac{\Gamma\sigma_i}{(1+\Gamma\sigma_i)},
\label{P_mu_el}
\end{equation}
where  $T^{*}=k_{B}T\epsilon\sigma_{i}/q^{2}$.

Finally, expressions for the  pressure  and for the partial  chemical potentials in the AMSA can be written as follows:
\begin{eqnarray}
\beta P^{AMSA}&=&\beta P^{ID}+\beta P^{HS}+\beta P^{MAL}+\beta P^{EL},
\label{P_amsa} \\
\nu_{i}^{AMSA}&=&\nu_{i}^{ID}+\nu^{HS}_{i}+\nu^{MAL}_{i}+\nu^{EL}_{i},
\label{nu_i_amsa} \\
\nu_{s}^{AMSA}&=&\nu_{s}^{ID}+\nu^{HS}_{s}.
\label{nu_s_amsa}
\end{eqnarray}

One can obtain the pressure and the chemical potentials in the MSA  by neglecting in   (\ref{P_amsa})-(\ref{nu_i_amsa}) the addends
connected with associations (those with the superindex ``MAL'').

Now, some remarks are in order. The results obtained in the AMSA depend on the definition of the ion pair and hence on the 
association constant $K^{0}$. As in \cite{HolPatPat17},
we choose $K^{0}$ in the form proposed by Olaussen and Stell \cite{Olaussen91}. We refer to it as $K^{0}_{OS}$. It was shown that
$K^{0}_{OS}$ provides the best agreement of the RPM critical parameters  with simulations \cite{Jiang02}.
It should
be indicated that $ K^{0}_{OS}\approx 12K^{0}_{Eb}$~\cite{Raineri00}, where $K^{0}_{Eb}$ is the association constant introduced by
Ebeling \cite{Ebeling68}
\begin{eqnarray*}
K^{0}_{Eb}(T)&=&\frac{2}{3}\pi\sigma_{i}^{3}\{b^{3}[E_{i}(b)-E_{i}(-b)]-b(e^{b}-e^{-b}) \\
&&
-(2+b^{2})(e^{b}+e^{-b})+ 6b^{2}+4\}.
\end{eqnarray*}
Although Ebeling's definition of the ion-association constant provides the correction of equation of state 
to the second
ionic-virial coefficient, it does not produce  good values for the critical temperature and critical density of the
RPM \cite{HolPatPat17,Jiang02}.

\section{Results and Discussion}
In this section we present results for the phase diagrams of the equisized RPM-HS mixture obtained from three theories, i.e.,
the WCA approximation, the MSA, and the AMSA.  
Coexistence curves are calculated at subcritical temperatures using the conditions of two-phase equilibrium
\begin{eqnarray}
 \nu_{i}(\rho^{\alpha},c^{\alpha},T)&=&\nu_{i}(\rho^{\beta},c^{\beta},T), \label{nui_eq} \\ 
 \nu_{s}(\rho^{\alpha},c^{\alpha},T)&=&\nu_{s}(\rho^{\beta},c^{\beta},T), \label{nus_eq} \\ 
 P(\rho^{\alpha},c^{\alpha},T)&=&P(\rho^{\beta},c^{\beta},T), 
 \label{P_eq}
\end{eqnarray}
 where $\rho^{\alpha(\beta)}$ is the total number density ($\rho=\rho_{i}+\rho_{s}$) in phase $\alpha(\beta)$ and $c^{\alpha(\beta)}$ is the concentration in phase $\alpha(\beta)$ expressed in terms of the mole fraction of solvent molecules ($c=\rho_{s}/\rho$).
The phase diagrams are built by solving numerically a set of equations Eqs.~(\ref{nui_eq})-(\ref{P_eq}) with respect to the densities $\rho^{\alpha}$ and $\rho^{\beta}$ and one of the concentrations $c^{\alpha}$ when the second concentration $c^{\beta}$
is given. Therefore, a series of the densities and concentrations in phases $\alpha$ and $\beta$ are obtained at temperatures of wide range.
To solve the set of equations Eqs.~(\ref{nui_eq})-(\ref{P_eq}), the Newton-Raphson iterative procedure has been used with an accuracy $10^{-9}$.

We have also calculated the critical lines using the classical conditions for the critical point of a two-component mixture~\cite{rowlinson_swinton}
\begin{eqnarray}
 \left(\frac{\partial^{2} G}{\partial c^{2}}\right)_{P,T}=0, \qquad
 \left(\frac{\partial^{3} G}{\partial c^{3}}\right)_{P,T}=0, \qquad
 \left(\frac{\partial^{4} G}{\partial c^{4}}\right)_{P,T}>0,
 \label{stable_SPM}
 \end{eqnarray}
where $G=G(P,T,c)$ is the Gibbs free energy and $c$ is  the mole fraction of solvent molecules. The first equation determines the bounding curve 
for material stability of the system \cite{rowlinson_swinton}. Eqs.~(\ref{stable_SPM}) are applicable equally to the vapour-liquid and liquid-liquid critical points of a binary mixture. These equations can be presented in terms of the derivatives of Helmholtz free energy $F$ with respect to the volume $V$ and the concentration $c$ (see Appendix~A). 

\subsection{Phase diagram of the RPM}
\begin{figure}[htb]
\begin{center}
\includegraphics [height=0.4\textwidth]{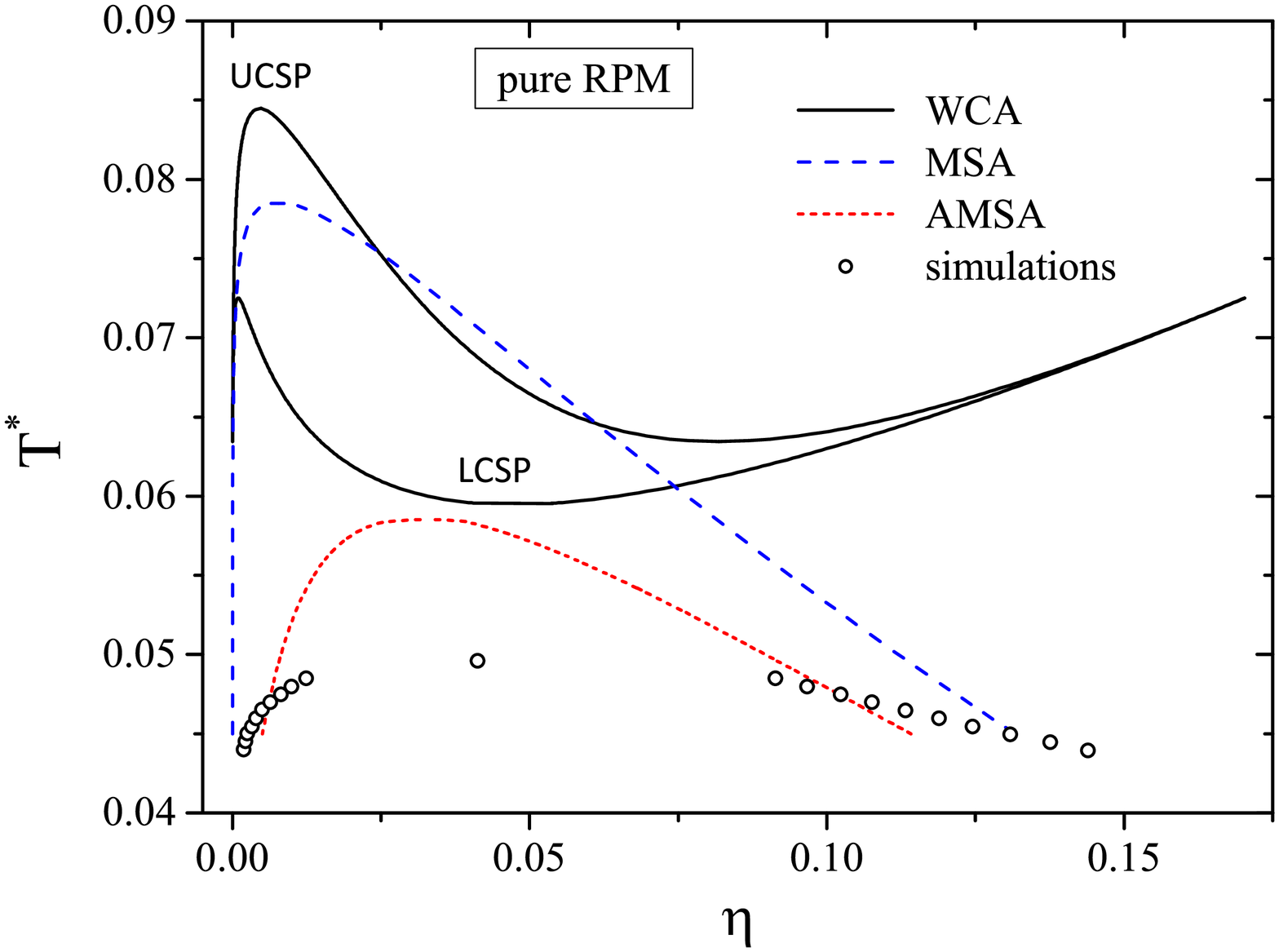}
\caption{Phase diagram of a pure RPM fluid obtained theoretically in three approximations, i.e., the WCA approximation, the MSA, and the AMSA, and by simulations. Lines correspond to the theoretical results and symbols denote the results of MC simulations \cite{yan-pablo:01}. In the AMSA, the association constant $K^{0}=K^{0}_{OS}$ (see the text for details).}
\label{Fig1}
\end{center}
\end{figure}

\begin{figure}[htb]
\begin{center}
\includegraphics [height=0.4\textwidth]{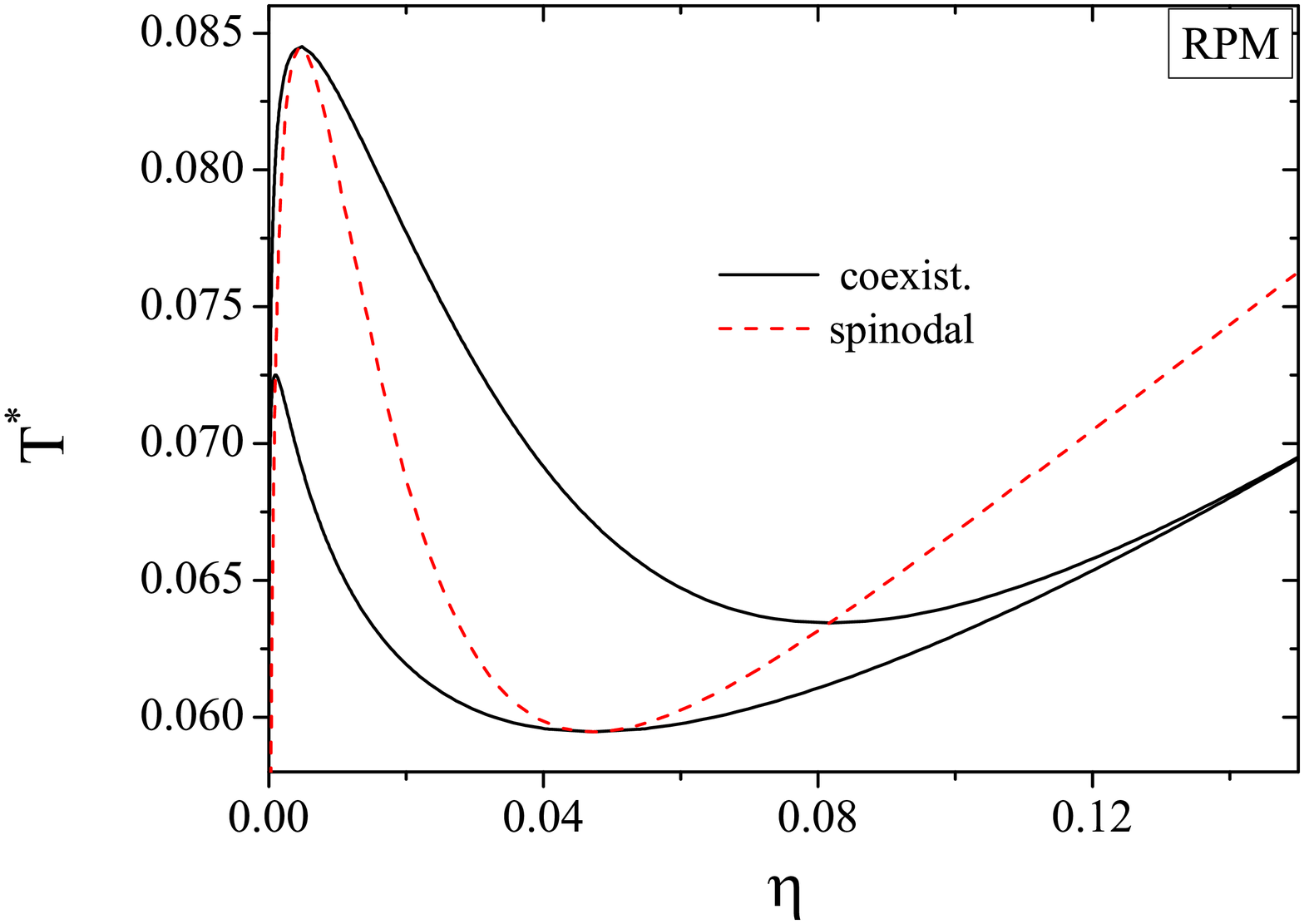}
\caption{Phase diagram of a pure RPM fluid obtained in the WCA approximation. 
The coexistence curves are presented by solid lines and the spinodal is presented by a dashed line.
}
\label{Fig2}
\end{center}
\end{figure}

We start with the phase diagram of the pure RPM fluid. In this case, Eqs.~(\ref{nui_eq})-(\ref{P_eq}) reduce to the two equations, 
for  $\nu_{i}$ and for $P$, under conditions $\rho^{\alpha(\beta)}=\rho_{i}^{\alpha(\beta)}$ and $c^{\alpha(\beta)}=0$. 
In Fig.~\ref{Fig1}, we show the coexistence curves obtained in the WCA, MSA and AMSA approximations and presented in terms of $T^{*}$-$\eta$
coordinates where $T^{*}=k_{B}T\epsilon\sigma_{i}/q^{2}$ is the reduced temperature and $\eta=\frac{\pi}{6}\rho_{i}\sigma_{i}^{3}$ is the packing fraction of ions.
As expected, all these approximations lead to an essential overestimation of the critical temperature and an underestimation of the critical density
in comparison with the computer simulations (see maxima on the curves). 
However, the vapour-liquid coexistence curve obtained in the AMSA approximation is noticeably better than in the WCA and MSA.
The MSA yields slightly better results for the critical parameters than the WCA, especially for the critical density (see Table~1).
At the same time, the diagram calculated in the WCA demonstrates a richer phase behavior of the RPM fluid.  
Apart from  the vapour-liquid coexistence, the upper curve of the WCA phase diagram also contains the branch appearing at higher densities 
(for $\eta>0.08$), which may indicate another type of phase behaviour induced by the charge ordering.
In addition, a lower region of the phase coexistence is found at low temperatures ($T^{*}<0.65$) with 
the corresponding lower critical point at $T_{L,c}^{*}=0.0595$ and $\eta_{L,c}=0.0472$ (see a minimum on the phase diagram). 
Therefore, the phase behaviour obtained in the WCA qualitatively differs from that described by the MSA and AMSA approaches.
We note that the regions where a coexistence of more than two different phases are found by the WCA theory (at higher densities or 
at lower temperatures) 
appear due to a multiplicity of solutions for Eqs.~(\ref{nui_eq}) and (\ref{P_eq}). It means that at some temperatures 
we have found more than two coexisting densities, which correspond to the same pressure and chemical potential.
However, it does not mean that all of the obtained phases can be considered as stable ones.
In order to check our results for the phase stability we have analysed the spinodal curve calculated 
for the RPM fluid using the WCA approximation. In Fig.~\ref{Fig2}, one can observe that the above-mentioned regions 
are located lower than the spinodal curve, hence we show that they are unstable.

Also the lower critical point is analysed using the conditions to be held for a stable critical point
\begin{equation}
 \left(\frac{\partial^{2} F}{\partial V^{2}}\right)_{T}=0, \qquad
 \left(\frac{\partial^{3} F}{\partial V^{3}}\right)_{T}=0, \qquad
 \left(\frac{\partial^{4} F}{\partial V^{4}}\right)_{T}>0.
 \label{stable_RPM}
\end{equation}
We have found that at $T^{*}=T_{L,c}^{*}$ and $\eta=\eta_{L,c}$, the first two equations of (\ref{stable_RPM}) are satisfied, while 
the inequality is not satisfied. Thus, the lower critical point obtained in the WCA approximation is unstable as well. 

\begin{table}[htbp]
\caption{The reduced  vapour-liquid critical parameters, $T_{c}^{*}$ and  $\rho_{c}^{*}$, for the restricted primitive model (RPM) obtained in
the theory and by simulations.}
\vspace{3mm}
\begin{tabular}{cccc}
\hline \hline\hspace{5mm}  Theory\hspace{6mm} &\hspace{6mm}
$T_{c}^{*}$\hspace{6mm} &\hspace{5mm}
$\rho_{c}^{*}$\hspace{8mm}
 &\hspace{5mm}
Ref.\hspace{8mm}
\\
\hline
WCA appr. & $0.08446$ & $0.0088$ &   \cite{patsahan_ion,Cai-Mol1}\\
MSA & $0.07858$ &$0.01449$&    \cite{Cai-Mol1}  \\
AMSA&      0.0587      &    0.0590    &    \cite{Jiang02}    \\
simulation&    0.0492   & 0.073       &    \cite{yan-pablo:01}    \\
\hline \hline
\end{tabular}
\end{table}

\subsection{Phase diagrams of the RPM-HS mixture}
The RPM-HS mixture undergoes the phase separation between a low-ionic-concentration phase  and a 
high-ionic-concentration phase.  Here, we address the phase diagrams of the RPM-HS mixture with $\sigma_{+}=\sigma_{-}=\sigma_{s}=\sigma$.

\begin{figure}[htb]
	\begin{center}
		\includegraphics[clip,width=0.47\textwidth,angle=0]{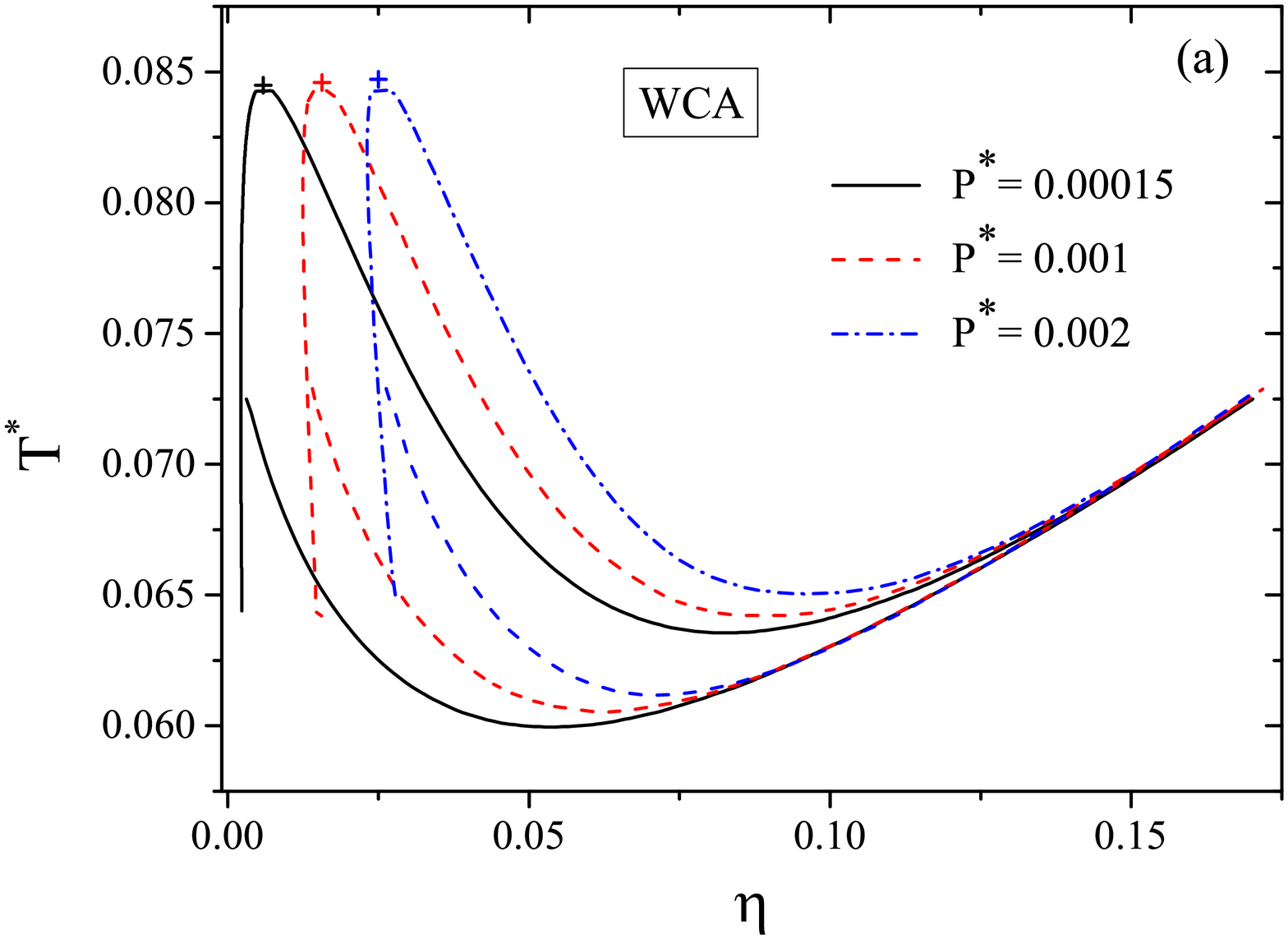}
		\includegraphics[clip,width=0.47\textwidth,angle=0]{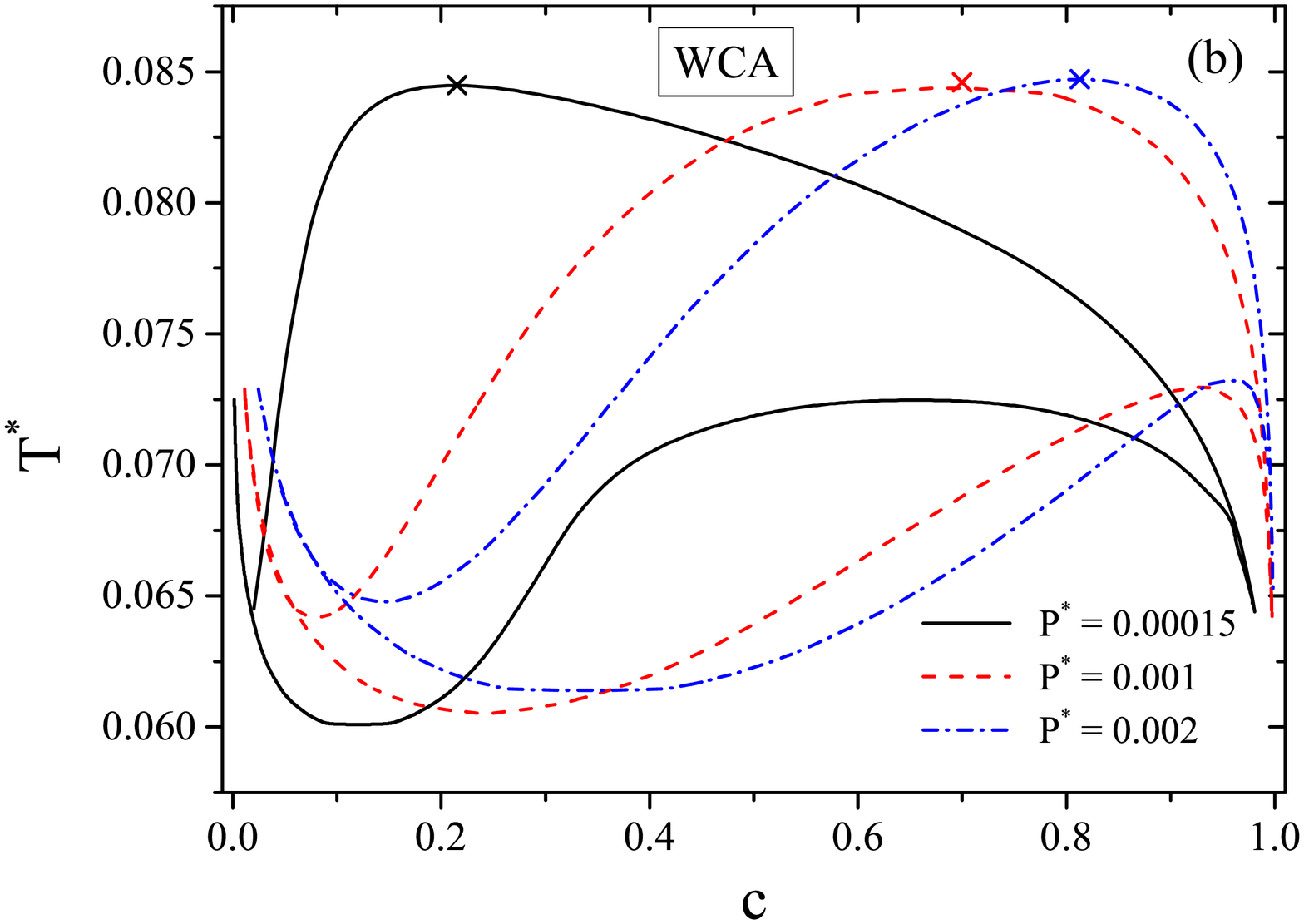}
		 \caption{
		Coexistence curves of the RPM-HS mixture in  $T^{*}$-$\eta$ (a) and $T^{*}$-$c$ (b) planes at constant reduced pressures in the WCA approximation. $T^{*}$, $\eta$, and $c$ are defined in Eq.~(\ref{notation}), and $P^{*}=P\varepsilon\sigma^{4}/q^{2}$.
	}
\label{Fig3}
	\end{center}
\end{figure}

First, we focus on the WCA approximation. In this case,  
the expressions for the partial chemical potentials, $\nu_{i}$ and $\nu_{s}$, and for the pressure $P$ are given 
by Eqs.~(\ref{nui_RPA_1})-(\ref{P_RPA_1})  taking into account Eq.~(\ref{wca_coul}). 
 In Figs.~\ref{Fig3}~(a)-(b), we show the 
coexistence curves which are calculated at fixed pressures   using Eqs.~(\ref{nui_eq})-(\ref{P_eq}).
The phase diagrams are presented in the $T^{*}$-$\eta$  and $T^{*}$-$c$  planes, where
\begin{eqnarray}
T^{*}=k_{B}T\epsilon\sigma/q^{2},
\qquad
\eta=\eta_{i}+\eta_{s}=\pi\rho\sigma^{3}/6, \qquad
c=\rho_{s}/\rho.
\label{notation}
\end{eqnarray}
The selected pressures are higher than  the critical pressure of the RPM in the WCA approximation ($P_{c}^{*}=P_{c}\varepsilon\sigma^{4}/q^{2}=7.85\times 10^{-5}$). 
It is seen in Fig.~\ref{Fig3}~(a) that the ($T^{*}$,$\eta$)-diagrams are of the shape similar to the 
phase diagram of the RPM obtained in the same approximation, i.e., they exhibit both the upper and the lower critical points.
It is also observed that an increase of pressure shifts the coexistence region towards higher total number densities.
The upper critical temperature is not affected essentially by the pressure, though a tendency to a small increase is noticed.
On the other hand, an increase of the lower critical temperature is more significant.
 For ($T^{*}$,$c$)-diagrams in Fig.~\ref{Fig3}~(b), one can see the closed miscibility loops and the lower 
 critical solution points (LCSP), the existence of which was also observed experimentally in \cite{Dittmar-Schroer} for ionic solutions in a non-polar solvent (the solution of $N_{4444}Br$ in toluene). 
However, it was shown in \cite{Dittmar-Schroer} that the phase transition related to the LCSP is located in a metastable region. 
Using the equations from Appendix~A, we have calculated positions of the critical points and have checked for their stability. 
We have found that, like for the RPM fluid, the lower critical point of the RPM-HS mixture is unstable.
The upper critical point is stable and moves toward higher concentrations of the solvent when the pressure increases. 
It should be noted that a solvent rich phase in Fig.~\ref{Fig3}~(b) corresponds to a lower-density branch of the coexistence curve in Fig.~\ref{Fig3}~(a) and vice versa.

\begin{figure}[ht]
	\begin{center}
		\includegraphics[clip,width=0.32\textwidth,angle=0]{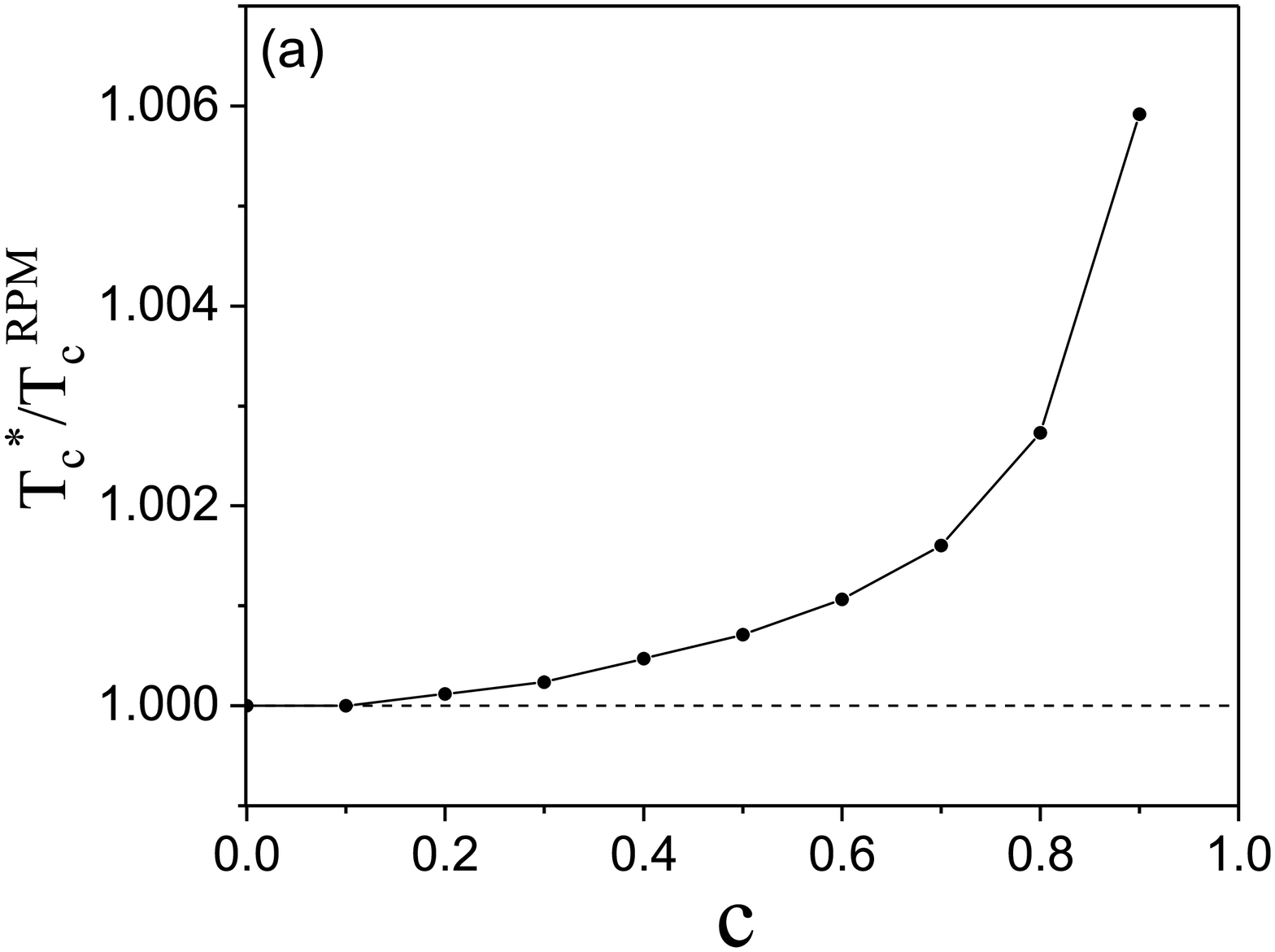}
		\includegraphics[clip,width=0.32\textwidth,angle=0]{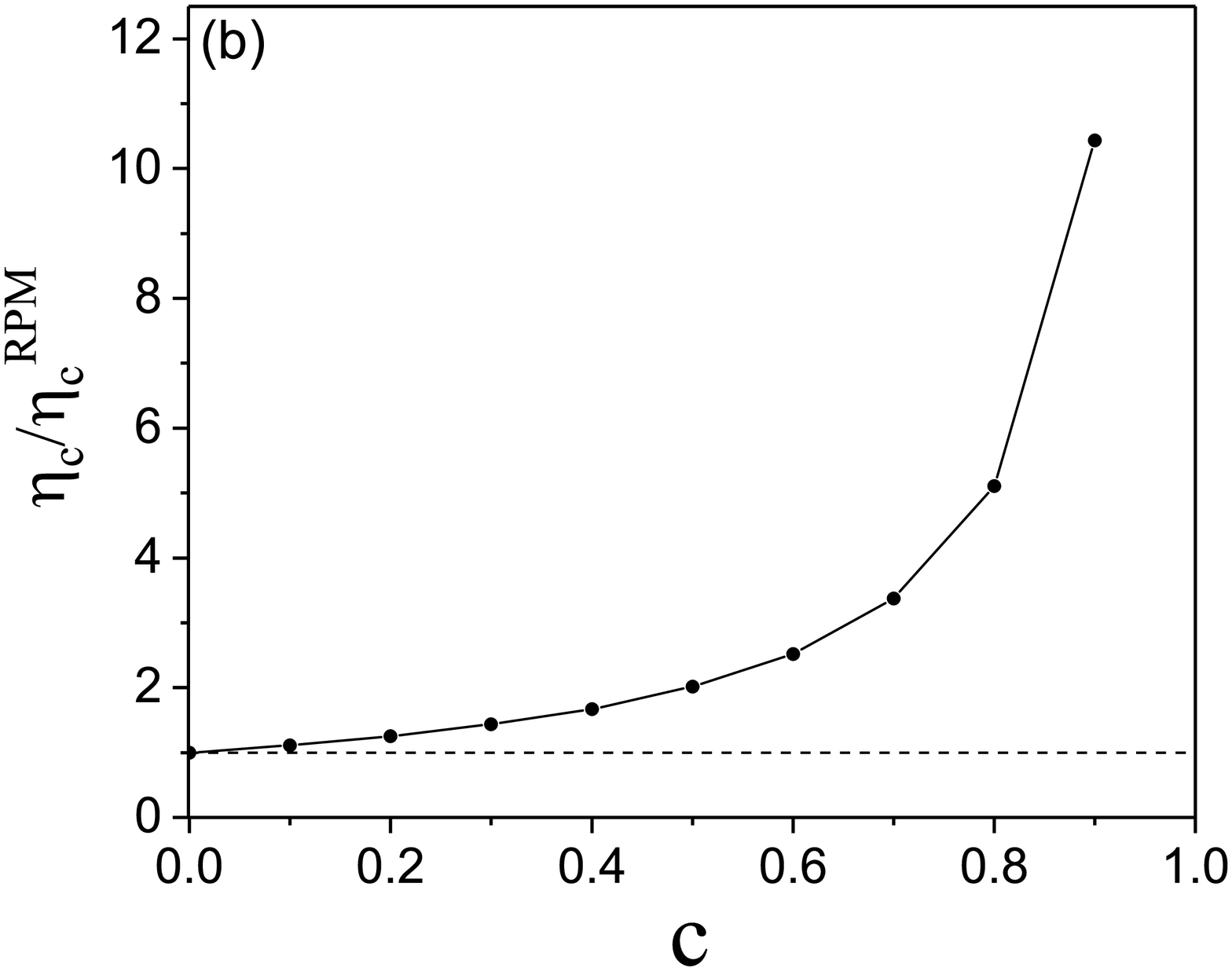} 
		\includegraphics[clip,width=0.32\textwidth,angle=0]{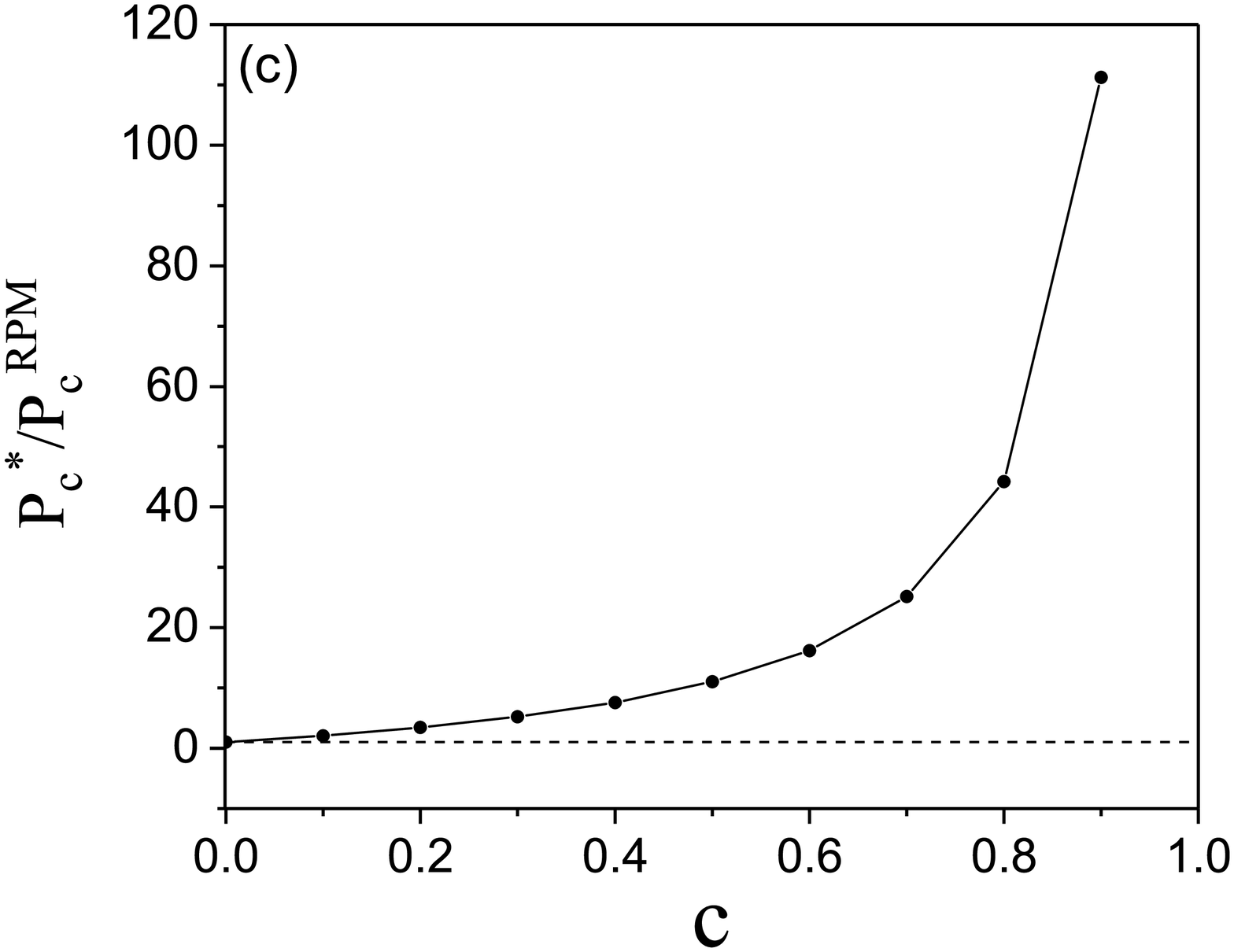} 
		\caption{
			The RPM-HS mixture: critical temperature $T_{c}^{*}/T_{c}^{RPM}$ (a), critical packing fraction 
			$\eta_{c}/\eta_{c}^{RPM}$ (b), 
			and critical pressure $P_{c}^{*}/P_{c}^{RPM}$ (c)  depending on the solvent concentration $c$ 
			in the WCA approximation. 
			$T_{c}^{RPM}$, $\eta_{c}^{RPM}$, and $P_{c}^{RPM}$ are the critical temperature,  the critical packing fraction, and 
			the critical pressure of a pure RPM fluid, respectively. $T^{*}$, $\eta$, and $c$ are defined in Eq.~(\ref{notation}), and $P^{*}=P\varepsilon\sigma^{4}/q^{2}$.
		}
		\label{Fig4}
	\end{center}
\end{figure}

We have analysed the dependencies of the upper critical point parameters on the solvent concentration $c$ taking the RPM critical point 
as a reference (Figs.~\ref{Fig4}~(a)--(c)).
One can see that an increase of the solvent concentration leads to a small increase of the critical temperature (Fig.~\ref{Fig4}~(a)) and to a 
significant increase of both the critical total packing fraction (Fig.~\ref{Fig4}~(b)) and the critical pressure (Fig.~\ref{Fig4}~(c)).

\begin{figure}[ht]
	\begin{center}
		\includegraphics[clip,width=0.47\textwidth,angle=0]{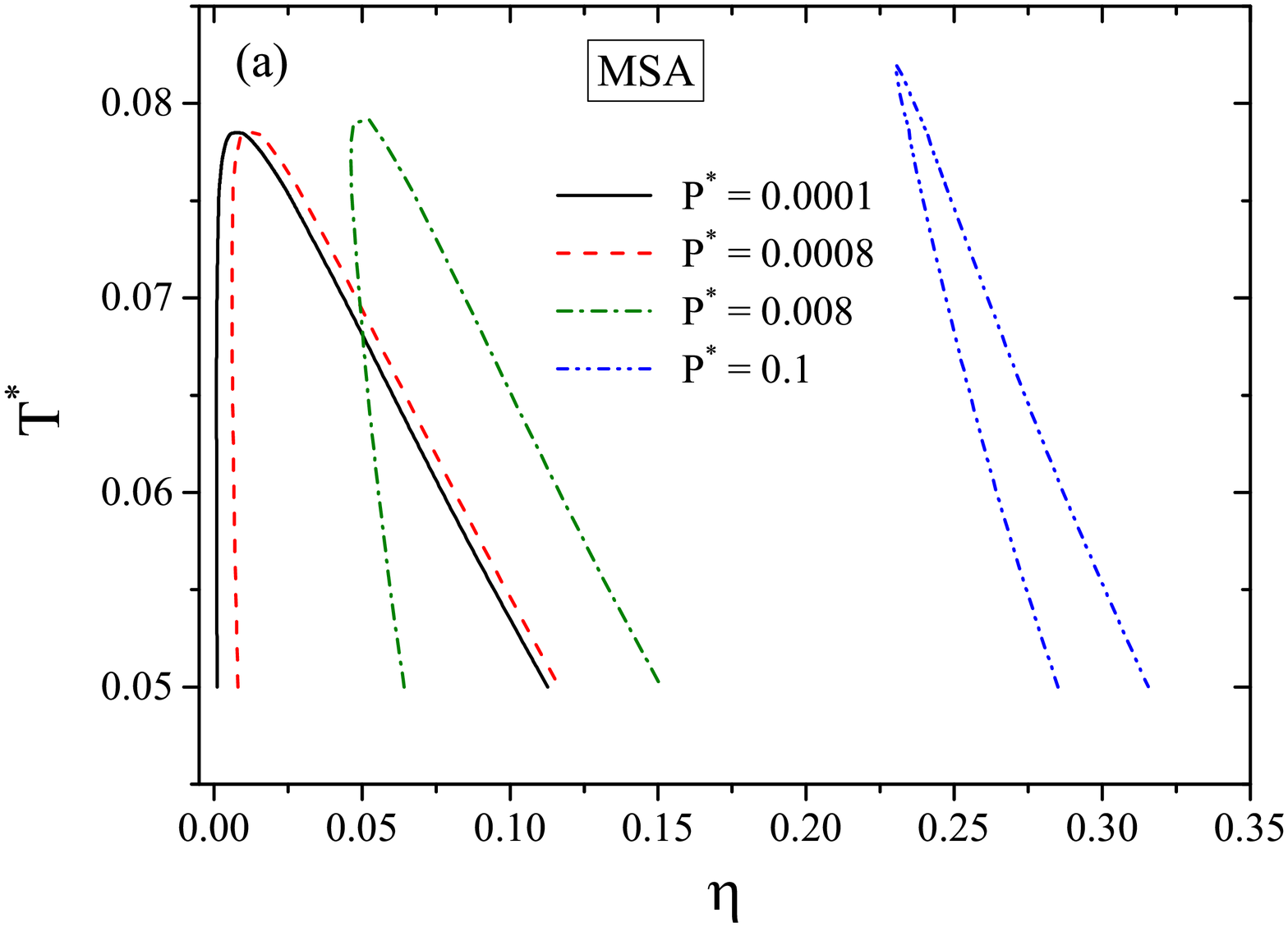}
		\includegraphics[clip,width=0.47\textwidth,angle=0]{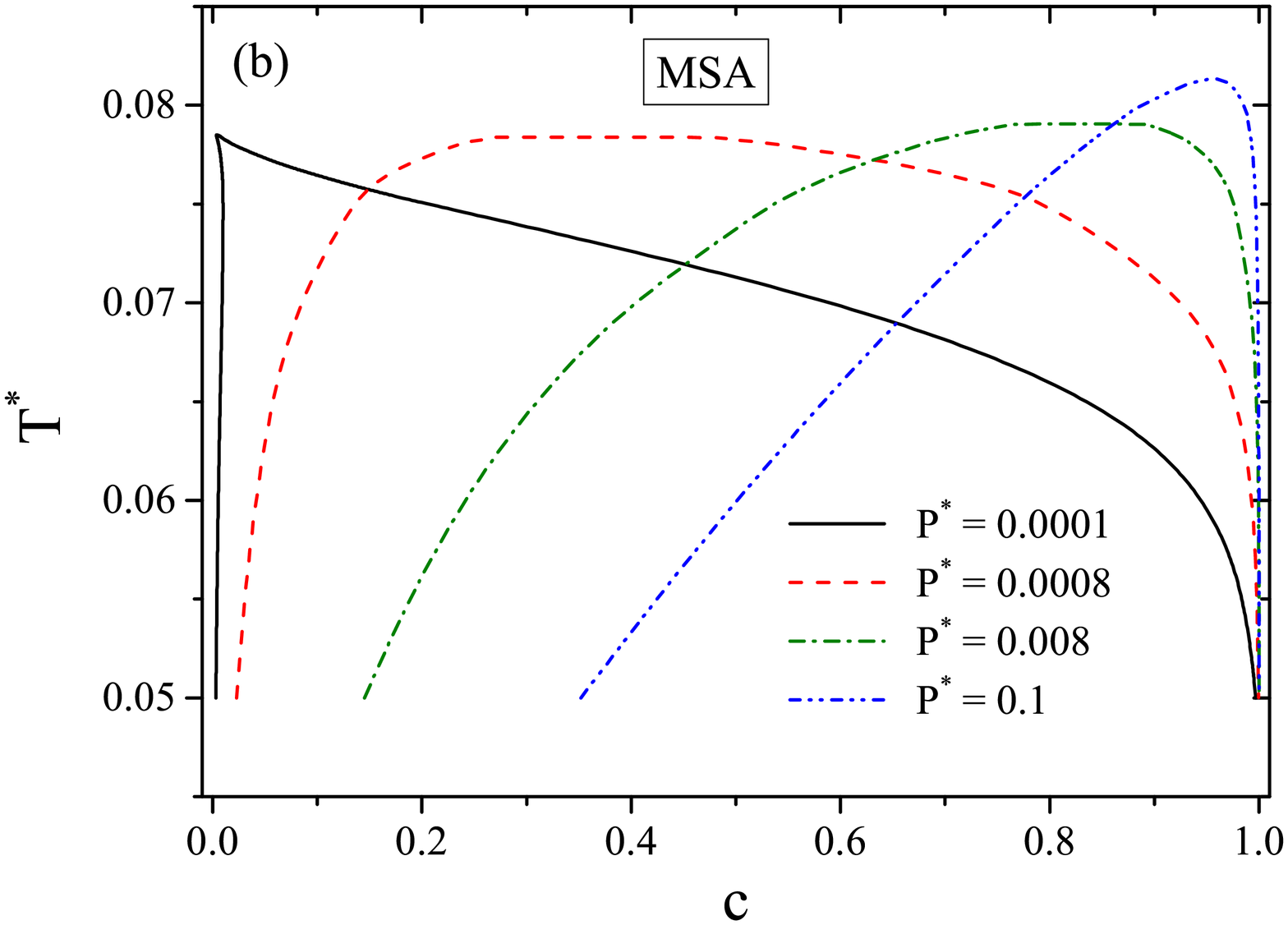} 
		\caption{Coexistence curves of the RPM-HS mixture in  $T^{*}$-$\eta$ (a) and $T^{*}$-$c$ (b) planes at constant reduced pressures in the MSA. $T^{*}$, $\eta$, and $c$ are defined in Eq.~(\ref{notation}), and $P^{*}=P\varepsilon\sigma^{4}/q^{2}$.
		}
		\label{Fig5}
	\end{center}
\end{figure}

Next, we have calculated the phase diagrams in the MSA. In this case, explicit expressions for the partial chemical potentials and pressure are obtained 
using Eqs.~(\ref{nui_RPA_1})-(\ref{P_RPA_1}) together with (\ref{reg-MSA1})-(\ref{reg-MSA2}). Figs.~\ref{Fig5}~(a)-(b) show the 
coexistence curves in the $T^{*}$-$\eta$ and $T^{*}$-$c$ planes obtained from Eqs.~(\ref{nui_eq})-(\ref{P_eq}). 
The phase diagrams are presented at pressures above the MSA critical pressure for the RPM that is $P^{*}=9.64\times 10^{-5}$. 
The selected pressures are the same as in \cite{Kenkare_SPM} (see Fig.~4 in \cite{Kenkare_SPM}) where the MSA results for the RPM-HS mixture are presented.  
A visual comparison made between our and the above-mentioned results indicates their quantitative agreement in the region 
of temperatures  $T_{c}^{*}>T^{*}>0.07$ for which  the coexistence is found in  \cite{Kenkare_SPM}. In our case, at each value of pressure,  
the phase coexistence is found for a wider region of temperatures, i.e., for $T_{c}^{*}>T^{*}>0.05$ which, in turn, 
corresponds to wider regions of packing fractions and concentrations. Nevertheless, neither phase transition related to the LCSP 
nor closed miscibility loop is observed in the MSA. For all of the considered pressures, we have obtained only the upper critical points.
For this type of phase transition, the obtained results qualitatively agree with the results of the WCA approximation. 
Quantitatively, the MSA produces a slightly lower critical temperature and a higher critical density than the corresponding critical parameters 
in the WCA approximation.

\begin{figure}[htb]
	\begin{center}
		\includegraphics[clip,width=0.47\textwidth,angle=0]{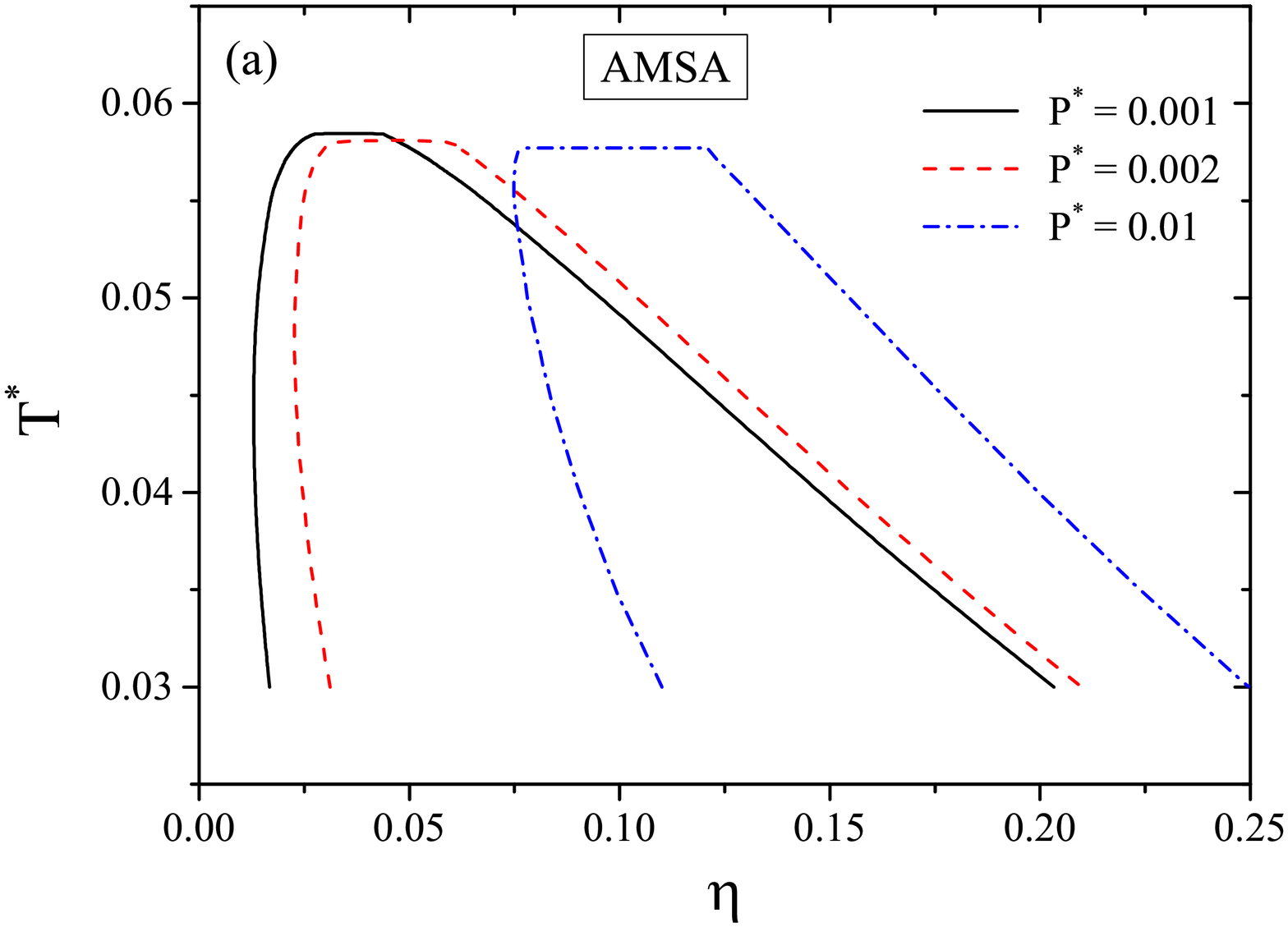}
		\includegraphics[clip,width=0.47\textwidth,angle=0]{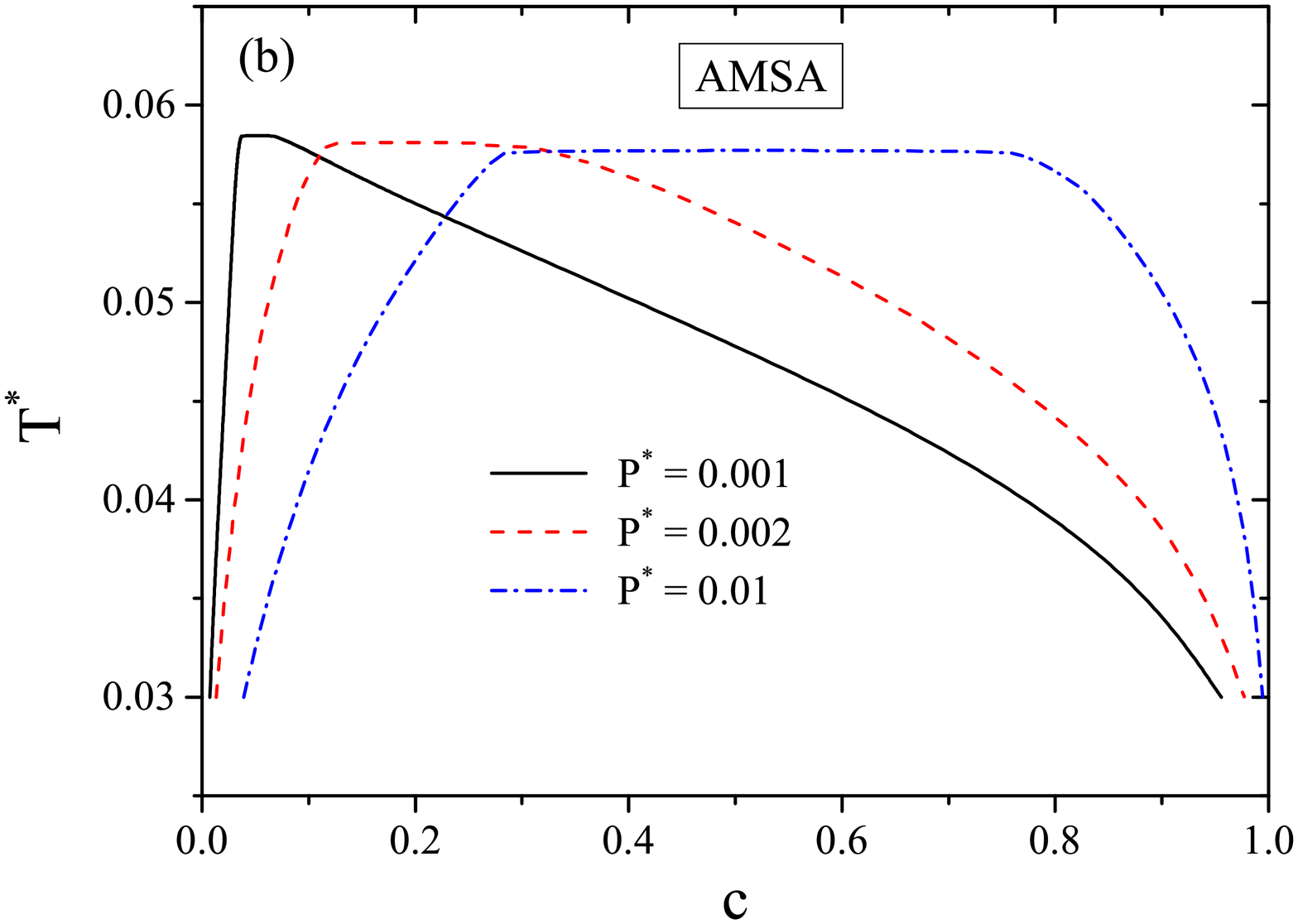}
		 \caption{Coexistence curves of the RPM-HS mixture presented in  $T^{*}$-$\eta$ (a) and $T^{*}$-$c$ (b) planes at constant reduced pressures 
		 in the AMSA. $T^{*}$, $\eta$, and $c$ are defined in Eq.~(\ref{notation}), and $P^{*}=P\varepsilon\sigma^{4}/q^{2}$.
	 }
\label{Fig6}
	\end{center}
\end{figure}

\begin{figure}[ht]
	\begin{center}
		\includegraphics[clip,width=0.47\textwidth,angle=0]{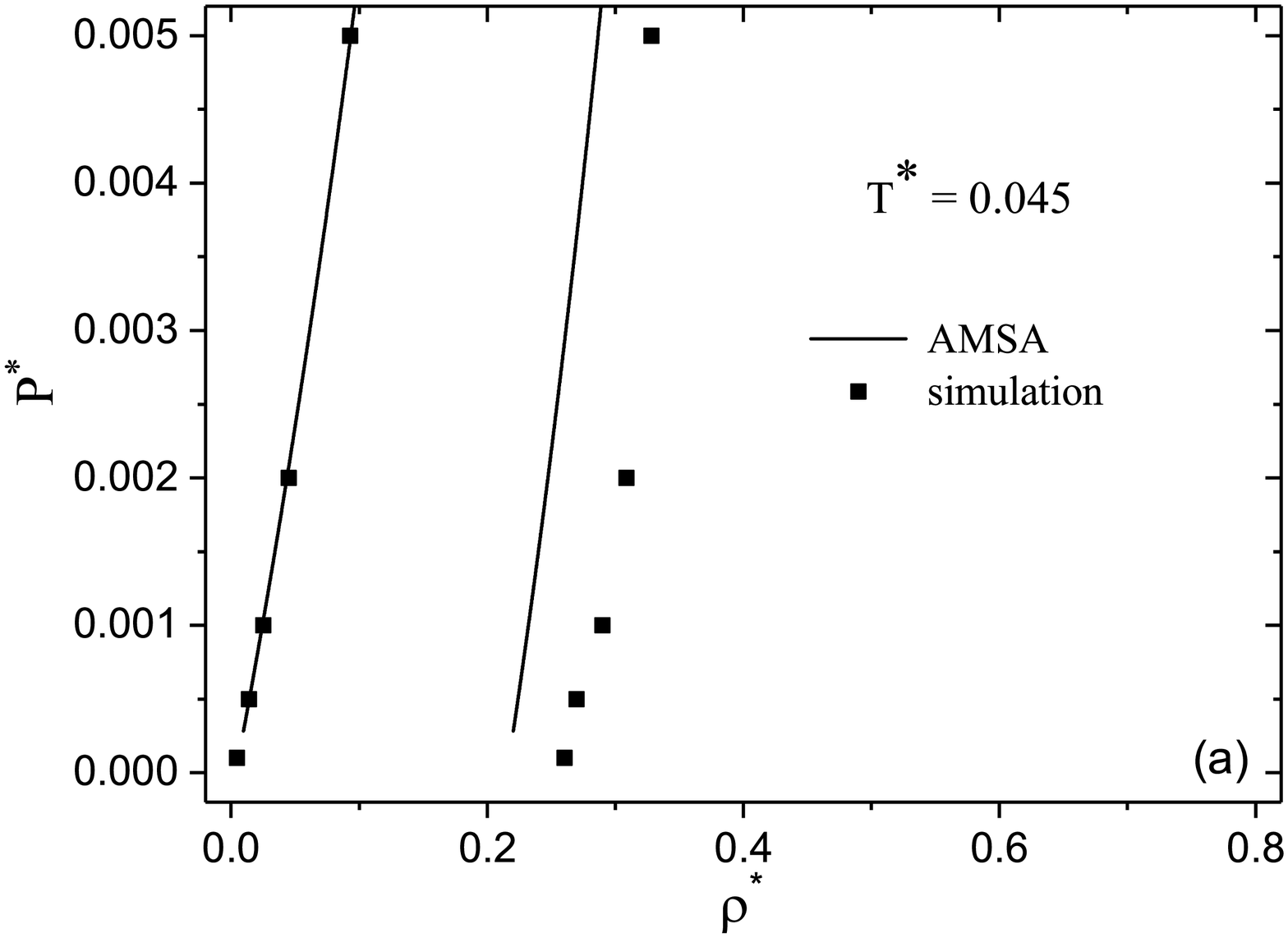}
		\includegraphics[clip,width=0.47\textwidth,angle=0]{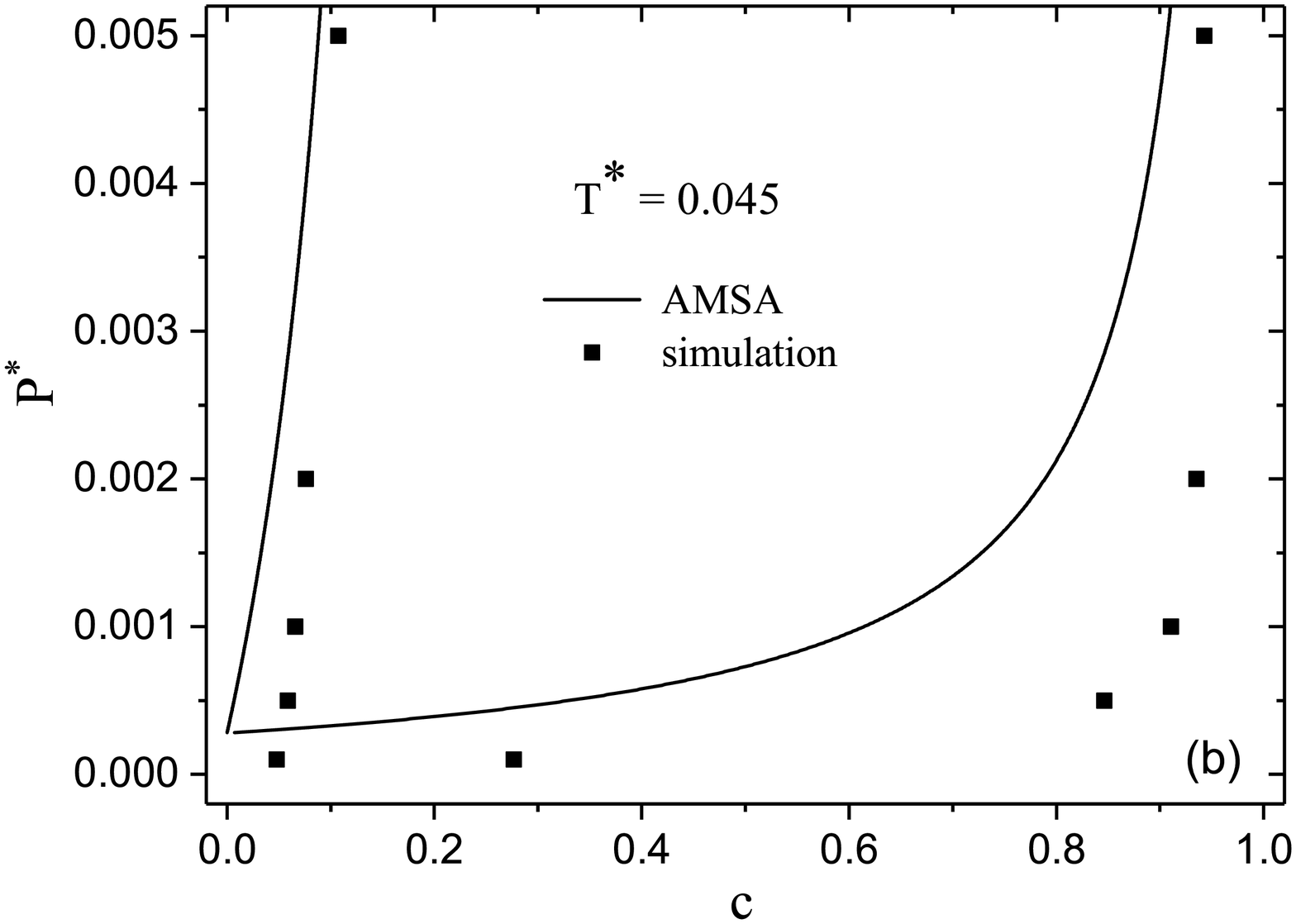}
		 \caption{
		Coexistence curves of the RPM-HS mixture presented in  $P^{*}$-$\rho^{*}$ (a) and $P^{*}$-$c$ (b) planes 
		 at constant temperature $T^{*}=0.045$. The lines are our AMSA results and the squares are the results 
		 of simulations 
		 \cite{Kristof_SPM}.  $P^{*}=P\varepsilon\sigma^{4}/q^{2}$, $\rho^{*}=\rho\sigma^{3}$, and $T^{*}$, $c$ are defined in Eq.~(\ref{notation}).
	}
\label{Fig7}
	\end{center}
\end{figure}

Finally, we present the coexistence curves calculated using the AMSA theory. Then, the partial chemical potentials and pressure are given by 
Eqs.~(\ref{P_mal})-(\ref{nu_s_amsa}) supplemented by the solution of Eq.~(\ref{Gamma_B}). 
The same as for the RPM, the association constant has been chosen to be  $K^{0}=12K_{Eb}^{0}$. 
The phase diagrams in the $T^{*}$-$\eta$ and $T^{*}$-$c$ planes at constant pressures are shown in Figs~\ref{Fig6}~(a),(b). All the 
selected values of pressure are higher than the critical pressure of the RPM ($P_{c}^{*}=7.44\times 10^{-4}$). It is worth noting 
that the critical pressure of the pure ionic system obtained in the AMSA is by an order higher than the corresponding pressure found from the MSA and WCA theories.
In general, the phase behaviour is qualitatively similar to that found in the MSA. 
The LCSP phase transitions have not been found in the AMSA either. Due to extreme flatness of the tops of coexisting curves obtained in the AMSA it is impossible to determine an exact 
location of the critical points. Although we can estimate the critical temperature with a rather good numerical accuracy, 
the critical densities and concentrations cannot be well defined. On the other hand, they can be estimated as  half the  sum of 
the densities  (concentrations) in coexisting phases observed at the highest possible temperature below the critical temperature.
The AMSA theory yields substantially lower values of the critical temperature and higher values of the critical total number density
than the MSA. 
Moreover, the critical temperatures obtained from the AMSA theory fall within the region predicted by simulations \cite{Shelley_SPM}. 
It is worth noting that in \cite{Shelley_SPM}, only the estimates of 
upper and lower bounds of the critical temperature are presented. Since the AMSA provides the prediction of critical points 
close to the critical points
obtained from simulations, we compare our AMSA results with the coexisting curves obtained in simulations by Krist\'{o}f et~al. \cite{Kristof_SPM} (see Figs.\ref{Fig7}~(a)-(b)) at the temperature $T^{*}=0.45$. As it is seen, the AMSA theory yields a reasonable agreement 
with the simulation results.

\section{Conclusions}
We have applied  well-known theoretical approaches, i.e., the CV theory, the MSA, and the AMSA to the study of fluid-fluid phase 
equilibria in the model of ionic solutions 
that takes into account  the presence of the solvent explicitly. Using the CV theory, we have found free energy, 
pressure and  partial chemical potentials in the RPA for a rather general model that takes into consideration   solvent-solvent  and  
solvent-ion interactions beyond the hard core. In this paper,
we have focused on RPM-HS mixture consisting of  oppositely charged hard spheres and neutral hard spheres. Within the RPA, 
we arrive at free energy in  the WCA approximation  by exploiting the WCA regularization of the Coulomb potential inside the hard core \cite{Cai-Mol1} and at  free energy in the ORPA/MSA by using the optimized regularization of the Coulomb potential \cite{waisman_lebowitz}.  It should be emphasised that the WCA approximation and the AMSA theory are applied to the study of the phase behaviour  of  the RPM-HS mixture for the first time. Despite the fact that the RPM-HS mixture had been previously studied by the MSA  \cite{Kenkare_SPM},  in this paper we  calculated the MSA phase diagrams  for a wider region of thermodynamic parameters. Moreover, the calculation of the coexistence curves in the MSA  serves as a verification of our results.  

We have calculated the fluid-fluid coexistence curves using the equations of phase equilibrium. The three above-mentioned approximations 
produce qualitatively similar results for the phase diagrams, i.e., an increase of pressure   shifts a fluid-fluid coexistence region 
towards higher total number densities and towards higher solvent concentrations and at the same time  leads only to a small increase of 
the  critical temperature. It should be noted that  the equations of the phase equilibrium  yield  more complex phase diagrams  in  the 
WCA approximation  compared to the MSA and AMSA. In particular, they consist of upper and lower branches that form a closed loop with 
the LCSP.  However, the thermodynamic  analysis shows that a lower branch and the corresponding critical point are unstable. 
Remarkably,  the LCSP and a nearly  closed miscibility  loop  located in the metastable region were reported for an ionic solution
in a non-polar solvent (the solution of $N_{4444}Br$ in toluene) \cite{Dittmar-Schroer}.
A quantitative comparison of the results obtained in the three approximations indicates that critical temperatures provided by the AMSA
are lower than in the WCA and MSA, while critical total number densities are higher. As for a pure RPM, the AMSA approximation leads to the best agreement 
with simulation findings. However, this result is defined by the association constant, which in our study is chosen in the form proposed 
by Olaussen and Stell~\cite{Olaussen91}. 

The next steps towards a theoretical description of the fluid-fluid phase behaviour in ionic solutions are to include other 
details of solvent-solvent and solvent-ion interactions, i.e., attraction/repulsion interactions beyond the hard core as well as 
a size (shape) asymmetry of ions and ion/solvent species. This work is now in progress using an analytical expression for the RPA 
free energy derived in Sec.~2.  
Another important issue is to go beyond the RPA in the study of explicit solvent  models. This can be done within the framework of 
the CV theory. As it was shown in \cite{PatPat09,patsahan-mryglod-patsahan:06,Patsahan_Patsahan:10,PatMryg12}, the 
CV approach is capable of reproducing, at least qualitatively, the effects of size and charge asymmetry on the vapour-liquid phase 
behaviour of the PMs, the solvent-free models of ionic solutions. In addition, the CV theory and the AMSA approach can 
be extended to the explicit solvent models which takes into account a non-spherical shape of solvent  molecules.

\section*{Acknowledgement}
This project has received funding from the European Unions Horizon 2020 research and
innovation programme under the Marie Sk{\l}odowska-Curie grant agreement No 734276.

\appendix

\section{Equations for the critical points of a two-component mixture}

The first two equations entering (\ref{stable_SPM})  expressed  in terms of  derivatives of the Helmholtz free energy
are as follows \cite{rowlinson_swinton}:
\begin{equation*}
F_{2c}F_{2V}-F_{Vc}^{2}=0, 
\end{equation*}
\begin{equation*}
F_{3c}F_{2V}^{2}-3F_{V2c}F_{Vc}F_{2V}+3F_{2Vc}F_{Vc}^{2}-F_{3V}F_{2c}F_{Vc}=0,
\end{equation*}
where
\begin{equation*}
F_{nVmc}=\frac{\partial^{n+m} F}{\partial V^{n}\partial c^{m}}
\end{equation*}
and $V$ and $c$ are the system volume and the solvent concentration, respectively.

The inequality in (\ref{stable_SPM})  is found to be
\begin{eqnarray*}
&&F_{4c}F_{2V}^{3}-4F_{2V}^{2}F_{Vc}F_{V3c}+6F_{2V}F_{Vc}^{2}F_{2V2c}-4F_{Vc}^{3}F_{3Vc}+
F_{2c}F_{Vc}^{2}F_{4V} \nonumber \\
&&
-3F_{2V}^{2}F_{V2c}^{2} -12F_{Vc}^{2}F_{2Vc}^{2}-6F_{3V}F_{Vc}^{2}F_{V2c}+12F_{2V}F_{Vc}F_{2Vc}F_{V2c}\nonumber \\
&&
+12F_{2c}F_{Vc}F_{2Vc}F_{3V}-3F_{2c}^{2}F_{3V}^{2}>0.
\end{eqnarray*}

For the RPM-HS mixture, we obtain  in the WCA approximation
\begin{eqnarray*}
\beta F_{2V}&=&\frac{N}{V^{2}}\left(a_{0}-24\eta c^{2}i_{2}\right), 
\\
\beta F_{2c}&=&N\left(\tilde{a}_{1}-24\eta i_{2}\right),  
\\
\beta F_{Vc}&=&\frac{N}{V}24\eta c i_{2},  
\end{eqnarray*}
\begin{eqnarray*}
\beta F_{3V}&=&-\frac{N}{V^{3}}\left[a_{2}-72\eta c^{2}\left(i_{2}-16\eta c i_{3} \right) \right], 
\\
\beta F_{3c}&=&-N\left[a_{1}-2(24\eta)^{2}i_{3} \right],  
\\
\beta F_{2Vc}&=&-\frac{N}{V^{2}}48\eta c\left(i_{2}-24\eta c i_{3} \right), 
\\
\beta F_{V2c}&=&\frac{N}{V}24\eta\left(i_{2}-48\eta c i_{3} \right), 
\end{eqnarray*}
\begin{eqnarray*}
\beta F_{4V}&=&-\frac{N}{V^{4}}\left\{a_{3}+2a_{2}-144\eta c^{2}\left[2i_{2}-64\eta c i_{3}+(24\eta c)^{2}i_{4}\right]\right\}, 
\\
\beta F_{4c}&=&-N\left[a_{11}-6(24\eta)^{3}i_{4} \right],  
\\
\beta F_{3Vc}&=&\frac{N}{V^{3}}144\eta c\left[i_{2}-48\eta c i_{3}+(24\eta c)^{2}i_{4}\right], 
\\
\beta F_{V3c}&=&-\frac{N}{V}4(24\eta)^{2}\left(i_{3}-36\eta c i_{4} \right), 
\\
\beta F_{2V2c}&=&-\frac{N}{V^{2}}48\eta\left[i_{2}-96\eta c i_{3}+3(24\eta c)^{2}i_{4}\right]. 
\end{eqnarray*}
In the above equations, the following notations are used
\begin{eqnarray*}
a_{1}&=&\frac{1-2c}{c^{2}(1-c)^{2}}, \quad
\tilde{a}_{1}=\frac{1}{c(1-c)}, \quad
a_{11}=\frac{2(1-3c+3c^{2})}{c^{3}(1-c)^{3}}, 
\nonumber \\
a_{2}&=&\frac{2(1+7\eta+10\eta^{2}-10\eta^{3}+5\eta^{4}-\eta^{5})}{(1-\eta)^{5}}, \quad
a_{3}=\frac{24\eta(1+4\eta)}{(1-\eta)^{6}}, \label{a2}\\
\nonumber
\\
i_{n}&=&\frac{1}{\pi}\int_{0}^{\infty}\frac{y^{2}\sin(y)^{n}}{(T^{*}y^{3}+24\eta c\sin(y))^{n}}{\rm{d}}y, \qquad
y=k\sigma_{i}.
\label{In}	
\end{eqnarray*}	



\end{document}